\documentclass[conference]{IEEEtran}
\IEEEoverridecommandlockouts

\usepackage{graphicx}
\usepackage{textcomp}
\usepackage{xcolor}
\usepackage{booktabs}
\usepackage{float}
\usepackage{bm}
\usepackage{tabularx}
\usepackage{mathtools}
\usepackage{subcaption}
\usepackage{pdfpages}
\usepackage{multirow}
\usepackage{weiwAlgorithm}
\usepackage{amsthm}
\usepackage{amsmath}
\usepackage{amsfonts}
\usepackage{diagbox}
\usepackage{threeparttable}
\usepackage{adjustbox}
\usepackage{url}
\usepackage{tabu}
\usepackage{array}
\usepackage{hyperref}
\usepackage{cleveref}

\newcommand{\myparagraph}[1]{\vspace{0.5mm} \noindent \textbf{#1}.}
\newcommand{\myparagraphunder}[1]{\vspace{0.5mm} \noindent \underline{#1}.}
\newcommand{\myblue}[1]{{\color{black} #1}\xspace}

\newtheorem{theorem}{Theorem}        
\newtheorem{proposition}{Proposition}       

\newtheorem{definition}{Definition}[section]
\newtheorem{problem statement}{Problem Statement}[section]

\def\BibTeX{{\rm B\kern-.05em{\sc i\kern-.025em b}\kern-.08em
    T\kern-.1667em\lower.7ex\hbox{E}\kern-.125emX}}
    
\begin{document}
\pagestyle{plain}
\pagenumbering{arabic}

\title{Beyond Homophily: Community Search on Heterophilic Graphs} 

\author{\IEEEauthorblockN{Qing Sima}
\IEEEauthorblockA{\textit{University of New South Wales}\\
Sydney, Australia \\
q.sima@unsw.edu.au}
\and
\IEEEauthorblockN{Xiaoyang Wang}
\IEEEauthorblockA{\textit{University of New South Wales}\\
Sydney, Australia \\
xiaoyang.wang1@unsw.edu.au}
\and
\IEEEauthorblockN{Wenjie Zhang}
\IEEEauthorblockA{\textit{University of New South Wales}\\
Sydney, Australia \\
wenjie.zhang@unsw.edu.au}

}

\maketitle

\begin{abstract}
Community search aims to identify a refined set of nodes that are most relevant to a given query, supporting tasks ranging from fraud detection to recommendation.
Unlike homophilic graphs, many real-world networks are heterophilic, where edges predominantly connect dissimilar nodes.
Therefore, structural signals that once reflected smooth, low-frequency similarity now appear as sharp, high-frequency contrasts.
However, both classical algorithms (e.g., $k$-core, $k$-truss) and recent  ML-based models struggle to achieve effective community search on heterophilic graphs, where edge signs or semantics are generally unknown.
Algorithm-based methods often return communities with mixed class labels, while GNNs, built on homophily, smooth away meaningful signals and blur community boundaries.
Therefore, we propose Adaptive Community Search (AdaptCS), a lightweight framework featuring three key designs:
(i) an AdaptCS Encoder that disentangles multi-hop and multi-frequency signals, enabling the model to capture both smooth (homophilic) and contrastive (heterophilic) relations;
(ii) a memory-efficient low-rank optimization that removes the main computational bottleneck and ensures model scalability; and
(iii) an Adaptive Community Score (ACS) that guides online search by balancing embedding similarity and topological relations.
\myblue{Extensive experiments on both heterophilic and homophilic benchmarks demonstrate that AdaptCS outperforms the best-performing baseline by an average of 11\% in F1-score, retains robustness across heterophily levels, and achieves up to 2 orders of magnitude speedup over the strongest ML-based CS baselines.}
\end{abstract}

\section{Introduction}
Graphs serve as a crucial representation for complex relational data across various domains, such as social networks~\cite{tan2023higher,yiqiwang1}, citation networks~\cite{yiqiwang2,qdgnn}, and molecular structures~\cite{molecular_22_nips,rong2020self}.
Identifying a closely interrelated community based on a query node has been an important research topic within the database domain. 
Existing community search (CS) methods can be broadly categorized into algorithm-based and machine learning (ML)-based approaches. 
Algorithm-based approaches define communities based on structural cohesiveness, such as $k$-core~\cite{k-core, fang2020survey}, $k$-truss~\cite{k-truss,k-truss2017}, and $k$-clique~\cite{k-clique,k-clique2018}, identifying densely interconnected nodes through graph-theoretic measures and optimization criteria. 
In contrast, ML-based approaches are task-driven and leverage predictive models, explicitly defining communities using known node labels or types \cite{ics-gnn,qdgnn,coclep}. 
By harnessing learned embeddings, ML-based methods effectively identify nodes related to a query node, emphasizing semantic similarity and class consistency.

Traditional graph algorithms and neural models typically assume {homophily}, where connected nodes are likely to share similar attributes or belong to the same community~\cite{graphsage,gin_2018,gat_2017}.
However, real-world graphs often violate this assumption and exhibit {heterophily}, where edges predominantly link nodes with different labels or communities~\cite{heter_freq_filter_aaai_23,heter_luan2023graph,heter_unsupervised_aaai_23}.
For instance, in citation networks, nodes represent papers and edges denote citations: while within-domain citations reflect homophily, cross-domain ones are common and crucial for knowledge transfer.
For example, a biology paper may cite database works on subgraph matching to analyze protein structures, and medical papers often reference AI vision models for radiology image analysis.
Similar heterophilic patterns occur in many real scenarios: fraudsters tend to interact with legitimate users rather than other fraudsters~\cite{fraud_jianke_2023,linBP_2015_VLDB}, political discussions frequently occur across opposing viewpoints~\cite{h2gcn_2020beyond,li2023signed}, and proteins in molecular graphs consist of diverse amino acids with distinct properties~\cite{h2gcn_2020beyond,li2023homogcl}.
\myblue{Notably, the edge signs in these scenarios are typically not explicitly observed, which limits the applicability of signed-graph models that assume known positive/negative relations.}

\begin{figure}[t]
  \centering
  \includegraphics[width=\linewidth]{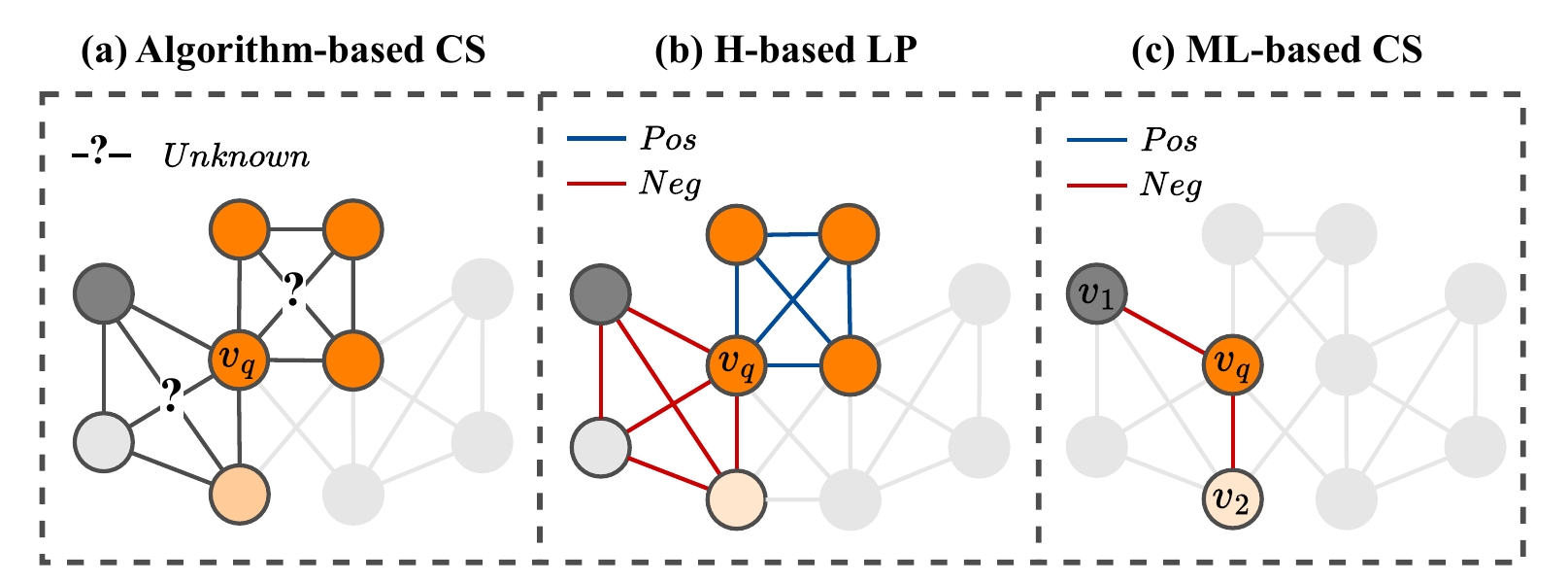}
        \vspace{-5mm}
  \caption{Limitations of three representative paradigms. 
  Node colors show communities; blue/red edges indicate implied homophilic/heterophilic links that are unobserved in practice.}
        \vspace{-1mm}
  \label{fig:three_graphs}
\end{figure}

\myparagraph{Existing solutions}
Although no prior work has directly addressed CS under heterophily, both the database and AI communities have studied graph heterophily on node classification tasks.
Building upon these insights, existing methods relevant to heterophilic CS can be broadly categorized into three paradigms:
(a) \textbf{Algorithm-based} approaches that rely purely on structural cohesiveness (e.g., $k$-core, $k$-truss, $k$-clique) and are not heterophily-aware;
(b) \textbf{Compatibility-matrix-based label propagation} methods~\cite{linBP_2015_VLDB,factorLP_2020_SIGMOD}, originally developed for heterophilic node classification, model cross-class relations through a compatibility matrix.
We extend this line of work to the CS by using the inferred compatibility patterns to guide community expansion around the query.
(c) \textbf{ML-based general solutions}, where heterophily-oriented GNN frameworks (e.g., FAGCN~\cite{FAGCN21}, ACM~\cite{ACM_2022}, ALT~\cite{ALT_KDD_2023}), originally proposed for node classification, are integrated as extensions into ML-based CS models to handle heterophily.

The above paradigms provide valuable insights yet exhibit fundamental limitations when applied to heterophilic CS.
\myparagraphunder{(a) Algorithm-based}
Classic structure-driven algorithms (e.g., $k$-core, $k$-truss, $k$-clique) identify dense subgraphs through purely graph-theoretic constraints.
However, without access to node labels or edge signs, these methods tend to produce mixed-label communities on heterophilic graphs, where edges frequently connect dissimilar nodes.
As depicted in \autoref{fig:three_graphs}(a), the algorithm uniformly aggregates nearby nodes based on structural density, often including irrelevant nodes.

\myparagraphunder{(b) Compatibility-matrix based label propagation ($H$-based)}
This line of methods~\cite{linBP_2015_VLDB,factorLP_2020_SIGMOD}, tackles heterophily in node classification by using a pre-defined or statistically learned compatibility matrix that governs how labels interact across edges.
The inferred matrix supports propagation across different classes without relying on the homophily assumption, and it can be adapted for community search by starting from a query node and expanding to other nodes predicted to have the same community.
However, the compatibility matrix $H$ is either predefined or globally optimized, remaining fixed across the graph and unable to adapt to local variations in heterophily.
As shown in \autoref{fig:three_graphs}(b), the query node $v_q$ connects to two distinct communities, where one is dominated by positive links and the other by negative links, exhibiting contrasting local homophily levels that a single global $H$ cannot accurately capture.
Consequently, such models often mispropagate signals across incompatible regions and fail to retrieve communities with heterophilic edge semantics.

\myparagraphunder{(c) ML-based CS (heterophilic extension)}
Recent  ML-based community search models (e.g., ICSGNN~\cite{ics-gnn}, QDGNN~\cite{qdgnn}, COCLEP~\cite{coclep}) are developed under an implicit homophily assumption and thus fail to generalize to heterophilic graphs.
Although general heterophily-aware GNNs such as FAGCN, ACM, and ALT can be applied as extensions to these models, their aggregation remains distance-agnostic: multi-hop signals are recursively blended across layers, giving rise to what we define as the Flip Effect (\autoref{def:flip-effect}) in multi-class settings.
As illustrated in \autoref{fig:three_graphs}(c), consider nodes $v_1$ and $v_2$ that are both heterophilic neighbors of the query node $v_q$.
Because the model mixes messages across hops, the two-step path $v_1\overset{-}{\longleftrightarrow}v_q\overset{-}{\longleftrightarrow}v_2$ introduces a false positive relation between $v_1$ and $v_2$ despite their different labels (different node colors).

\myparagraph{Challenges}
Despite distinct formulations, all three paradigms share intrinsic bottlenecks on heterophilic graphs:

\myparagraphunder{Unknown edge semantics}
Real-world heterophilic graphs rarely provide explicit polarity or semantics for edges, yet effective community search requires inferring which connections are “positive’’ or “negative’’ without explicit edge signs.

\myparagraphunder{Multi-hop inconsistency (Flip Effect)}
Aggregating multi-hop messages without explicitly disentangling distance information causes semantic inversion along even-hop paths, which leads to false positive relations between nodes of different classes.

\myparagraphunder{Lack of adaptivity}
Heterophilic graphs typically contain mixed homophilic and heterophilic regions; robust models must adapt dynamically to each graph and query, balancing topological and semantic consistency.

\myparagraph{Our solutions} 
To tackle these challenges, we propose \textbf{Adapt}ive \textbf{C}ommunity \textbf{S}earch (AdaptCS), a two-phase framework consisting of graph encoding and online search.
In the encoding phase, the AdaptCS encoder applies adaptive masking to extract exact‐$k$ hop neighborhoods without overlap, ensuring each channel contains only information from nodes at a fixed distance.
This distance-aware decomposition individually processes information from different hops, thereby avoiding the flip effect.
The resulting hop‐specific features are further processed by a frequency‐aware filter that splits them into low‐pass (smooth, homophilic) and high‐pass (non‐smooth, heterophilic) components.  
Finally, a lightweight two‐dimensional channel mixer fuses hop and frequency channels into compact node embeddings, preserving both local detail and long‐range context. 
To further improve scalability, AdaptCS employs a memory-efficient low-rank optimization that computes all hop-specific features in latent space, avoiding explicit high-order adjacency materialization and eliminating the main efficiency bottleneck. 
In the online search phase, AdaptCS incorporates:  
(i) a Signed Community Search (SCS), which utilizes the learned embeddings to construct a positive graph, and perform CS accordingly; and 
(ii) an Adaptive Community Score (ACS) that dynamically balances embedding‐based similarity and topological relations according to the graph’s approximated homophily ratio.
Under high homophily, ACS places greater weight on connectivity, whereas under low homophily, it relies more on embedding signals.
The primary contributions of this paper are summarized as follows:

\begin{itemize}
    \item 
          To the best of our knowledge, this work is the first to tackle the community search problem in heterophilic graphs.
          We propose a hop-distinct aggregation to handle varying heterophily levels while mitigating flip effects.
    \item 
          We propose an Adaptive Community Score (ACS) that maintains robust community search performance under both homophilic and heterophilic graph structures.
    \item 
          We propose a low-rank approximation optimization that removes the major efficiency bottleneck, enabling our model to scale to graphs with hundreds of millions of edges on a single GPU without memory overflow.
    \item 
          Experiments on real-world graphs demonstrate consistent gains in community search accuracy, robustness, and computational efficiency over state-of-the-art baselines.

\end{itemize}

\section{Related Work} \label{related work} 
\myparagraph{Algorithm-based community search}
Traditional community search algorithms utilize various cohesiveness metrics to identify communities within graphs. Metrics such as $k$-core \cite{k-core,tan2023higher}, $k$-truss \cite{k-truss,k-truss2017}, and $k$-clique \cite{k-clique,k-clique2018} have been employed to efficiently detect subgraphs that meet predefined structural criteria. For instance, the $k$-core metric identifies subgraphs where each node has at least $k$ connections within the subgraph, ensuring a level of internal connectivity. $k$-truss focuses on the presence of triangles, identifying subgraphs where each edge participates in at least $k-2$ triangles, thus capturing a higher-order cohesiveness. These methods have been extended to attributed graphs, incorporating node attributes alongside structural considerations to identify communities of nodes sharing similar characteristics \cite{attributed,attributed1}.

\myparagraph{Machine learning-based community search}
The advent of GNNs has introduced flexible and expressive models for community search, capable of balancing contributions from both topological structures and node attributes. 
Models like ICSGNN \cite{ics-gnn} leverage GNNs to capture similarities between nodes by combining content and structural features. 
This approach allows for the interactive and iterative discovery of target communities, guided by user feedback. 
Similarly, QDGNN \cite{qdgnn} employs an offline setting, training on a fixed dataset and extending to attributed community search by adopting an attribute encoder to identify groups of nodes with specific attributes. 
Other models, such as ALICE \cite{jianwei_sig} and COCLEP \cite{coclep}, incorporate advanced techniques like cross-attention encoders and contrastive learning to enhance the expressiveness and efficiency of community search in attributed graphs.
More recently, SMN~\cite{smn_25} and CommunityDF~\cite{chen2025communitydf} further advance this line of research: SMN proposes a general solution for overlapping community search via subspace embedding, while CommunityDF introduces a generative diffusion-based framework that iteratively refines query-centered subgraphs through contrastive learning and dynamic thresholding.

\myparagraph{Learning on heterophilic graphs}
Another research line extends classical label propagation to heterophilic settings through a learnable compatibility matrix that models class interactions across edges.
LinBP~\cite{linBP_2015_VLDB} reformulates belief propagation into a linear system supporting both homophilic and heterophilic relations via a global matrix $H$, while FactorLP~\cite{factorLP_2020_SIGMOD} further learns $H$ from labeled data through matrix factorization.
These approaches preserve the interpretability and efficiency of propagation but remain globally linear, unable to adapt to local heterophily or leverage node attributes.

Beyond propagation models, recent GNN-based methods aim to enhance robustness under heterophily.
Approaches such as Geom-GCN~\cite{geom-gcn}, MixHop~\cite{Mixhop}, and GPRGNN~\cite{GRPGNN_ICLR_2021} expand or reweight message passing to capture high-order and flexible dependencies, while FAGCN~\cite{FAGCN21} introduces signed edge weighting to model cross-class relations.
General frameworks like ACM~\cite{ACM_2022} and ALT~\cite{ALT_KDD_2023} further extend these ideas by adaptively combining multi-channel propagation or optimizing graph structure, enabling GNNs to handle graph heterophily.

\section{Preliminaries} \label{Preliminaries}
Let $G=({V,E})$ be an undirected graph with a set ${V}$ of nodes and a set ${E}$ of edges. 
Let $n =  |{V}|$ and $m = |{E}|$ be the number of nodes and edges. 
Given a node $u \in {V}$, ${N}_u=\{v|(u, v) \in {E} \}$ is the neighbor set of $u$. 
The adjacency matrix of $G$ is denoted as ${A} \in \{0, 1\}^{n\times n}$, where ${A}_{i,j} = 1 \text{, if } (v_i, v_j) \in {E}$, otherwise ${A}_{i,j} = 0$. 
${X} \in \mathbb{R}^{n \times d}$ is the set of node features, where $d$ is the dimension of the feature, and ${x}_i$ represents the node features of $v_i$. 
We use ${Z} \in \mathbb{R}^{n \times c}$ to denote the label matrix, whose $i$-th row is the one-hot encoding of the label of $v_i$ and $c$ is the dimension of the label.

\subsection{Problem Definition}

The key characteristic of a heterophilic graph is that for a given node \( u \), a majority of its neighbors belong to different community labels, i.e., $P(v \in N_u \mid z_v \neq z_u) > P(v \in N_u \mid z_v = z_u)$, where \( N_u \) denotes the neighborhood of node \( u \). 
This means that information propagation in such graphs fundamentally differs from homophily settings, where neighboring nodes will more likely belong to the same class.
To quantify the level of heterophily in a graph, we use the edge homophily metric \( h_{edge}(G) \), defined as:
\begin{equation} \label{eq:heter_metric}
h_{edge}(G) = \frac{| \{ (u,v) \in E \mid z_u = z_v \} |}{|E|}.
\end{equation}
A heterophilic graph exhibits a low \( h_{edge}(G) \), meaning that most edges connect nodes with different community labels.
Following the existing definition of the CS in ML-based models~\cite{ics-gnn,qdgnn,coclep,smn_25}, we define the CS as below:

\begin{problem statement}[Community Search (CS)]\label{def:cs}
Given a graph \( G = (V, E) \), a query node \( q \), and a target community size \( \mathcal{K} \), 
the CS task aims to identify a node set \( V_c \subseteq V \) that is semantically consistent with the query node \( q \). 
Formally,
\(
V_c \subseteq \{\, v \in V \mid z_v = z_q \,\}, \quad |V_c| = \mathcal{K},
\)
where \( z_v \) and \( z_q \)  denote the community label of node \( v \) and the query node \( q \).
\end{problem statement}

\begin{figure*}[t]
  \centering
  \includegraphics[width=\textwidth]{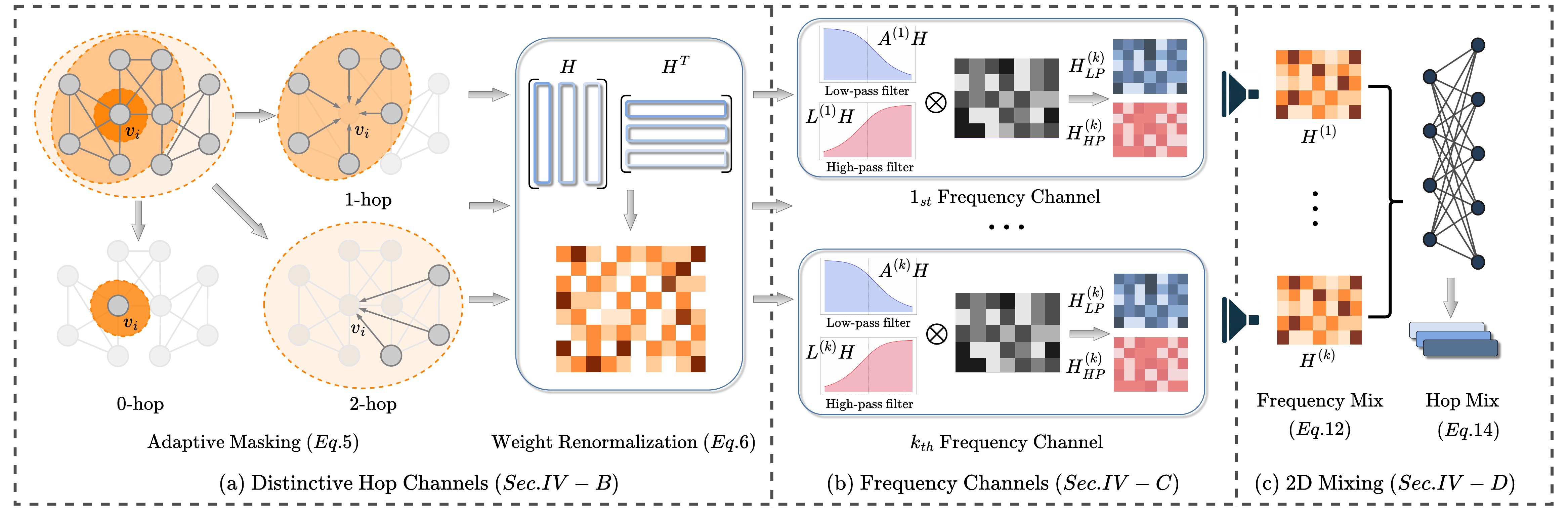}
        \vspace{-6mm}
  \caption{\myblue{AdaptCS Encoder Frameworks}}
        \vspace{-4mm}
  \label{fig:framework}
        \vspace{-3mm}
\end{figure*}

\begin{figure}[t]
  \centering
  \includegraphics[width=\linewidth]{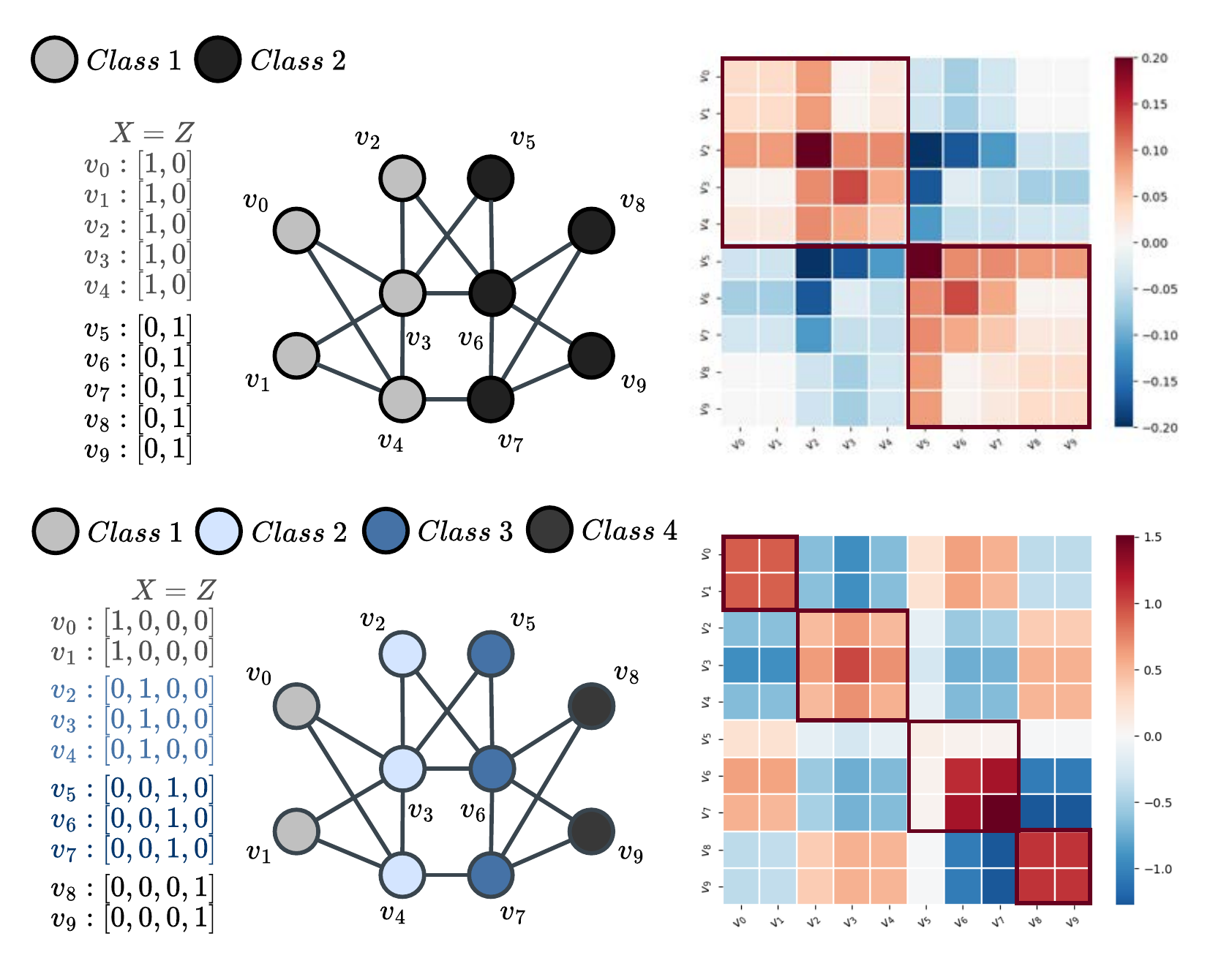}
        \vspace{-6mm}
  \caption{Flip Effects}
        \vspace{-1mm}
  \label{fig:case_study}
\end{figure}

\subsection{Graph Laplacian}
\myblue{In heterophilic graphs, low-pass aggregation with normalized adjacency can over-mix features from different labels.
To preserve heterophily-informative signals, our encoder additionally uses a high-pass channel.
The graph Laplacian is the standard high-pass operator; thus we briefly review it below and use it to construct our high-frequency propagation.}

\myblue{The graph Laplacian is defined as \(L = D - A\), where $D$ is the degree matrix, which is symmetric and positive semidefinite}. 
Variations include the degree normalized Laplacian (a symmetric matrix), defined as \(\tilde{L}_{\mathrm{sym}} = D^{-1/2}\,L\,D^{-1/2} = I - D^{-1/2}{A}D^{-1/2}\), and random walk normalized Laplacian (non-symmetric), defined as \(\tilde{L}_{\mathrm{rw}} = D^{-1}\,L\, = I - D^{-1}{A}\).
In GCN framework \cite{gcn2016}, the low-pass output is computed as 
\begin{equation} \label{eq:gcn}
Y = \mathrm{softmax}\Bigl(\hat{A}_{\mathrm{sym}}\,\cdot\mathrm{ReLU}(\hat{A}_{\mathrm{sym}}\,X\,W_0)\,W_1\Bigr),
\end{equation}
where \(W_0 \in \mathbb{R}^{f \times f_1}\) and \(W_1 \in \mathbb{R}^{f_1 \times o}\) are learnable weight matrices. 
$\hat{A}_{\mathrm{sym}}$ is the low-pass filter, and can be replaced by the high-pass filter $\hat{L}_{\mathrm{sym}}$ to form a high-frequency channel.
The random walk renormalized matrix \(\hat{A}_{\mathrm{rw}} = \bar{D}^{-1}\bar{A}\) and its corresponding Laplacian \(\hat{L}_{\mathrm{rw}} = I - \hat{A}_{\mathrm{rw}}\) serve as mean aggregators in spatial-based GNNs, sharing same eigenvalues as $\hat{A}_{\mathrm{sym}}$.
We use \(\hat{A}_{\mathrm{sym}}\) in this paper and denote it as \(\hat{A}\).

\subsection{Flip Effect}

\begin{definition}[Flip Effect]
\label{def:flip-effect}
Given a multi-label graph \( G = (V, E) \), let \( z_u \) denote the community label of node \( u \in V \). 
Suppose the graph is heterophilic, such that edges often connect nodes with different labels. 
The flip effect arises when aggregating information across multiple hops: 
if there exists a two-hop path \( u \overset{-}{\longleftrightarrow} w \overset{-}{\longleftrightarrow} v \) with \( z_u \ne z_w \), \( z_w \ne z_v \) and \( z_u \ne z_v \), 
standard aggregation may falsely infer \( z_u = z_v \), thereby ``flipping'' the underlying relationship.
\end{definition}

\autoref{fig:case_study} presents a toy example illustrating how high-frequency (Laplacian) aggregation behaves differently in binary and multi-class heterophilic graphs.
The left side shows the one-hot node embeddings ($X=Z$) and class labels, and the right side displays the similarity matrix $LXX^\top L^\top$ as a heatmap, where $L$ is the graph Laplacian filter.

\myparagraph{Binary case} 
When only two classes are present, intra-class node pairs (outlined blocks on the diagonal) exhibit strong positive similarity (red), while inter-class pairs are negative (blue). 
This means that Laplacian aggregation accurately separates nodes from the two given communities, assigning positive similarity only to nodes from the same class. 

\myparagraph{Multi-class case}  
With the four-class case, ideally, the red color should only appear on the diagonal (highlighted inter-class regions).
However, the same aggregation produces additional off-diagonal red blocks, notably between Class 1 and Class 3, and between Class 2 and Class 4.
These correspond to node pairs that are connected via an even number of negative edges (such as two-hop paths), causing their similarity to be incorrectly flipped from negative to positive.
Hence, Laplacian-based high-frequency filtering fails to distinguish communities in the multi-class setting, merging structurally distinct groups and demonstrating the "flip effect"—an inherent limitation of high-pass aggregation in general heterophilic graphs.
          


\section{Methodology Phase \romannumeral 1: Graph Encoding}\label{sec:AdaptCS}
\myblue{
Adaptive Community Search (AdaptCS) is a two-phase framework consisting of graph encoding and online search. During the graph encoding phase, AdaptCS learns community-oriented node representations. 
In the online search phase, the model leverages these learned representations to retrieve a meaningful query-centric community given an input query. 
This section presents the technical details of the AdaptCS encoder; the online search algorithm is described in Sec.~\ref{sec:online}.
}
\subsection{Encoder Frameworks}
\autoref{fig:framework} gives an end-to-end overview of our AdaptCS encoder. 
\myblue{
Starting from an input graph \(G=(V,E,X)\), the model framework disentangles hop distance (Distinctive-hop channels) and spectral frequency (Frequency channels), then fuses the resulting 2D signal by a channel mixer.}

\myparagraph{Stage 1: Distinctive-hop channels (Sec.\ref{sec:hop})}
\myblue{In \autoref{fig:framework}(a), given a node \(v_i\), we recursively conduct \(k\)-hop neighborhoods aggregation to construct distance-aware features. 
The $k$-th channel only aggregates information from neighbors with an exact shortest distance of $k$, which eliminates redundancy and noisy messages during convolution.  
We then apply weight renormalization by assigning edge-wise similarity scores via dot products between the embedding matrix $H \in \mathbb{R}^{n \times h}$, which rescales the aggregated messages and stabilizes magnitudes.
}

\myparagraph{Stage 2: Frequency channels (Sec.\ref{sec:freq})}  
\myblue{Frequency filters (low pass and high pass) process the hop-specific feature matrices to produce two complementary views: a smooth (homophilic) representation and a non-smooth (heterophilic) representation (\autoref{fig:framework}(b)).}
This separation allows each channel to specialize in either cohesive similarity or distinctive contrast without mutual interference.

\myparagraph{Stage 3: 2D-channel mixing (Sec.\ref{sec:mix})}  
\myblue{In \autoref{fig:framework}(c), we first use attention to fuse the low- and high-frequency channels within each hop, and then concatenate the fused representations across all hops. }
Next, a multi-layer perceptron (MLP) or a per-class attention bank is applied, producing a unified node embedding that integrates both local details and long-range context in a compact representation.

The resulting embeddings serve as the foundation for two online community search algorithms, SCS and ACS. 
Together, the three stages produce robust node embeddings that capture both local and global heterophilic structures, enabling accurate and scalable community search on heterophilic graphs.

\subsection{Distinctive Hop Channels (Distance-aware)}\label{sec:hop}
In this section, we propose distinctive hop channels to explicitly process signals from different hops of neighbors, designed for different homophily levels.
Distinctive hop channels address three limitations in traditional multi-hop aggregation approaches. 
First, this approach solves the flip effect, where signals from each hop are processed independently, handling the complex semantics between higher-order relations. 
Second, the distinctive hop adjacency matrices are shortest-distance aware, where each node in the \( k \)-th hop channel accurately represents edges with the shortest path distance of exactly \( k \). 
Third, this approach reduces redundancy and improves space efficiency since the sum of all distinctive hop adjacency matrices precisely reconstructs the cumulative adjacency \(\hat{A}^k\), avoiding repetitive edge representation.
\myblue{
To explicitly measure whether a representation preserves label-consistent similarity under such settings, we introduce the following notion.}
\myblue{
\begin{definition}[Heterophily-aware Node Distinguishability (HND)]
\label{def:HND}
Given a labeled graph $G=(V,E)$, let $z_u$ denote the class label of node $u\in V$.
Let $\{h_u\}_{u\in V}$ be the hidden state and define similarity
$S(u,v)=\mathrm{sim}(h_u,h_v)$.
In multi-class heterophily ($c>2$), node $u$ is heterophily-distinguishable if
\[
\mathbb{E}\!\left[S(u,v)\mid z_v=z_u\right]\ \ge\ \mathbb{E}\!\left[S(u,v)\mid z_v\neq z_u\right].
\]
We define $\mathrm{HND}(G)$ as the fraction of nodes $u$ for which the above condition is well-defined and holds.
\end{definition}

HND serves as a direct representation-level criterion: a higher value indicates that the learned embedding geometry remains consistent with labels despite heterophilic connectivity.
Our key idea is to make multi-hop aggregation distance-aware by constructing exact-$k$-hop channels, and then use a prototype bank to select which hop signals should dominate the final representation.
The following proposition summarizes this mechanism and its implications for HND.


\begin{proposition}[Distance-aware hop selection improves HND]
\label{prop:flip}
AdaptCS constructs exact-$k$-hop representations $\{h_u^{(k)}\}$ and fuses them via prototype-gated hop attention.
Under hop separability and mild gating margins, the fusion upweights the label-consistent hop, yielding embeddings that satisfy
the HND ordering in Definition~\ref{def:HND} and thus improve $\mathrm{HND}(G)$.
The formal guarantee is given in Theorem~\ref{thm:HND_DA}.
\end{proposition}

}

\myparagraph{Segregate multi-hop signals distinctively} We introduce two masking methods for achieving distinctive hop decomposition. 

\myparagraphunder{Hard Masking} This method strictly excludes any previously encountered connections in the next iteration:

\begin{equation}
\label{eq:hard-mask}
\hat{A}^{(k)}
= \mathrm{Mask}\Bigl(\hat{A}^k,\; \sum_{j=1}^{k-1} \hat{A}^{(j)}\Bigr),
\quad \hat{A}^{(0)} = I,\ \hat{A}^{(1)} = \hat{A},
\end{equation}

\noindent where the \(\mathrm{Mask}\) operator is defined as:
\begin{equation}
\mathrm{Mask}(X,Y)_{i,j} =
\begin{cases}
X_{i,j}, & \text{if } Y_{i,j}=0,\\
0,       & \text{otherwise}.
\end{cases}
\end{equation}

\noindent
Hard masking imposes a strict rule: any connection that has appeared at a previous hop is excluded from later hops.
While hard masking avoids redundant aggregation, it can also discard structural signals in densely-connected regions, where the same relations are repeatedly reinforced across multiple hops.
Hence, higher-hop propagation is forced to rely on newly emerging connections, which are typically less stable and more noise-sensitive.
To address this limitation, we introduce an adaptive masking scheme that selectively preserves edges whose multi-hop strength increases from the previous hop.

\myparagraphunder{Adaptive masking}
We define adaptive masking by
\begin{equation}
\label{eq:ReLU-mask}
\hat{A}^{(k)} \;=\; \mathrm{ReLU}\!\bigl(\hat{A}^{k} - \hat{A}^{k-1}\bigr),
\end{equation}
where $\mathrm{ReLU}$ zeros out the negative entries which retains two types of connections at hop $k\!\ge\!2$:
\begin{itemize}
\item New $k$-hop connections: entries that were zero at $(k{-}1)$ hops (similar to hard masking and expands reachability).
\item Strengthened existing connections: entries whose $k$-hop value strictly exceeds their $(k{-}1)$-hop value. These encode multi-hop reinforcement and are our focus below.
\end{itemize}

\myblue{
\myparagraph{Why the strengthened edges matter}
At $k{=}2$, strengthened existing edges $(u,v)\in E$ (i.e., $(\hat A^2-\hat A)_{uv}>0$) are exactly those receiving substantial two-hop support from common neighbors, hence they are highly embedded and triangle-rich.

\myparagraph{Why does this benefit the community search}
(i) Triangle-rich edges anchor locally dense subgraphs, concentrating similarity and making the endpoints more likely to fall into the same community; 
(ii) multiple common neighbors create more two-hop paths that average out label noise, bringing endpoints representation closer together even under heterophily (the neighbors’ labels may vary, but the shared-neighborhood signal remains coherent).
This yields a concrete, verifiable advantage at two hops, formalized below.

\begin{theorem}[Triangle-support lower bound for adaptively retained edges]
\label{thm:triangle_support_lb}
Let $\hat A=D^{-\frac12}(A+I)D^{-\frac12}$ and define $\hat A^{(2)}=\mathrm{ReLU}(\hat A^2-\hat A)$.
For any edge $(u,v)\in E$ with $(\hat A^2-\hat A)_{uv}>0$, write
\(
T(u,v)\coloneqq 1-\frac{1}{\hat d_u}-\frac{1}{\hat d_v}.
\)
The triangle support (or length of the common neighbor set $CN$) satisfies
\[
\mathrm{supp}(u,v)=|CN(u,v)|\;\ge\;\big\lfloor 3\,T(u,v)\big\rfloor+1.
\]
A detailed proof appears in Appendix A online \cite{adaptcs}.
\end{theorem}

\noindent{Beyond $k{=}2$.}
While Theorem~\ref{thm:triangle_support_lb} specifically certifies local triadic density at two hops, the same edges typically retain larger multi-hop mass at higher orders (via degree-normalized walk contributions), making such node pairs more {likely} to reside in the same locally dense subgraph. 

}

\myparagraphunder{Weight Renormalization} While adaptive masking effectively mitigates redundant edges and flip effects, it also causes the magnitude of $\hat{A}^{(k)}$ to decay rapidly as $k$ increases, which may lead to numerical instability and gradient vanishing.
To stabilize aggregation across hops, we introduce a renormalization step that rescales the masked adjacency using attention:
\begin{equation}\label{eq:local_weight}
\alpha_{ij}^{(k)}=\sigma\!\bigl((W\mathbf{h}_i)^{\!\top}(W\mathbf{h}_j)\bigr), \
\tilde{A}^{(k)}=RN(\hat{A}^{(k)}\odot\alpha^{(k)}),
\end{equation}
\myblue{where $W$ is learnable parameters, $\mathbf{h}_i$ denotes the node embedding for node $v_i$, $\sigma$ is the sigmoid function, $\alpha^{(k)} \in \mathbb{R}^{n \times n}$ is the weights, $RN$ is RowNorm which rescales each row to sum to one}.
This local normalization enables each node to prioritize semantically consistent neighbors and minimize the influence of noisy ones, thereby ensuring stable propagation and balanced gradient flow under heterophily.

\subsection{Frequency Channels} \label{sec:freq}
In the previous sections, we have constructed hop-distinct propagation operators $\{\hat{A}^{(k)}\}$, which naturally serve as {low-pass} channels that aggregate information from the exact-$k$ neighborhood. 
Following ACM~\cite{ACM_2022}, we further incorporate a high-pass channel to capture heterophily-informative components that low-pass aggregation may over-mix.
Unlike the original ACM, which recursively propagates features across layers, our framework leverages hop-distinct channels.

To avoid the prohibitive cost of explicitly constructing hop-wise Laplacian matrices, we exploit the identity that for any normalized adjacency $A$ with Laplacian $L=I-A$, the high-pass response can be computed efficiently as $LX=X-AX$. 

\myparagraph{Raw distinctive hop}  
For each hop $k$, with $A^{(k)}$ denoting the exact-$k$ adjacency, the hop-wise Laplacian operator is:
\begin{equation}
L^{(k)} = I - A^{(k)}.
\end{equation}

\myparagraph{Local weight renormalization}
When applying local weight renormalization, we first reweight the hop-$k$ operator by edge-wise attention as in Eq.~\eqref{eq:local_weight} and then propagate features.
\myblue{The $k$-hop features aggregated by a distinct low-pass filter $\tilde{A}^{(k)}$ is}
\begin{equation}
X^{(k)}_{\mathrm{LP}} \;=\; \tilde{A}^{(k)} X 
\;=\; RN\bigl(\hat{A}^{(k)} \odot \alpha^{(k)}\bigr)\, X .
\end{equation}
The reweighted high-pass branch is as follows:
\myblue{\begin{equation}
\tilde{A}^{(k)}_{\mathrm{HP}} \;=\; RN\bigl(\hat{A}^{(k)} \odot (1-\alpha^{(k)})\bigr),
\label{eq:hp_op}
\end{equation}
\begin{equation}
X^{(k)}_{\mathrm{HP}} \;=\; \tilde{A}^{(k)}_{\mathrm{HP}} X .
\label{eq:hp_feat}
\end{equation}}
In summary, our hop-distinct design computes both low-pass and high-pass responses per hop using shared operators, avoiding explicit Laplacian construction and additional memory overhead while enabling frequency-aware feature extraction.


\subsection{2D-Channel Mixing}\label{sec:mix}
After obtaining hop-distinct frequency features $\{X_{\mathrm{LP}}^{(k)}, X_{\mathrm{HP}}^{(k)}\}_{k=1}^{K}$ from the preprocessing pipeline, we propose a 2D-channel mixing stage to integrate both frequency and hop information, yielding robust node representations for downstream community search.

\myparagraph{Linear transformation}
\myblue{Before mixing, channel features are projected into a unified latent space via linear transformations, $H_{\mathrm{LP}}^{(k)}$ and $H_{\mathrm{HP}}^{(k)}$ are hidden state aggregated by low/high pass filter at the $k$-hop, calculate as follows:}
\begin{equation}
H_{\mathrm{LP}}^{(k)} = \mathrm{ReLU}(X_{\mathrm{LP}}^{(k)} W_{\mathrm{LP}}), \quad
H_{\mathrm{HP}}^{(k)} = \mathrm{ReLU}(X_{\mathrm{HP}}^{(k)} W_{\mathrm{HP}}),
\end{equation}
where $W_{\mathrm{LP}}, W_{\mathrm{HP}} \in \mathbb{R}^{d \times h}$ are learnable parameters and $d$ is the input size, $h$ is the hidden size. 

\myparagraph{Frequency channel mixing via attention}
For each hop $k$, the transformed embeddings $H_{\mathrm{LP}}^{(k)}$ and $H_{\mathrm{HP}}^{(k)}$ are fused using a node-wise adaptive attention mechanism. Specifically, we learn scalar attention weights for every node and channel:
\begin{equation}
(\alpha_{\mathrm{LP}}^{(k)},\,\alpha_{\mathrm{HP}}^{(k)}) = \mathrm{Attention}\big(H_{\mathrm{LP}}^{(k)},\, H_{\mathrm{HP}}^{(k)}\big),
\end{equation}
where the attention module outputs two weights that sum to one. The fused embedding at hop $k$ is then computed as
\begin{equation}
H^{(k)} = \alpha_{\mathrm{LP}}^{(k)} \odot H_{\mathrm{LP}}^{(k)} + \alpha_{\mathrm{HP}}^{(k)} \odot H_{\mathrm{HP}}^{(k)}.
\end{equation}

\myparagraph{Hop channel mixing via MLP}
To incorporate multi-hop information, we concatenate the frequency-fused embeddings from all hops along the feature dimension:
\begin{equation}
H_{\mathrm{concat}} = H^{(0)} \Vert H^{(1)} \Vert \dots \Vert H^{(K)},
\end{equation}
where $\Vert$ denotes concatenation. The unified node representation is then obtained by applying a linear layer or an MLP:
\begin{equation}
H_{\mathrm{final}} = \mathrm{ReLU}(H_{\mathrm{concat}} W_{\mathrm{hop}}),
\end{equation}
where $W_{\mathrm{hop}}$ is a learnable weight matrix for hop-wise fusion.

\myparagraph{Hop channel mixing via attention bank}
Another approach to fuse hop features is via a per-class attention bank. 
Given a learnable class bank $P \in \mathbb{R}^{d \times c}$, we first compute a node-wise class attention,
and derive the adaptive feature weight:
\begin{equation}
w^{\mathrm{class}} = \alpha^{\mathrm{class}} P^\top, \quad
\alpha^{\mathrm{class}} = \mathrm{softmax}(H^{(0)} P).
\end{equation}

We first apply $w^{\mathrm{class}}$ to scale features according to the specified class pattern, then utilize the same class specified weight for the computation of hop-wise weights $\alpha^{\mathrm{hop}}$:
\begin{equation}
H_{\mathrm{final}} = \sum_{k=1}^{K} \alpha^{\mathrm{hop}}_k \cdot \left( H^{(k)} \odot w^{\mathrm{class}} \right),
\end{equation}
where $\odot$ denotes element-wise multiplication. 
This procedure adaptively weights features and hop messages via attention.
Notably, the filter bank fusion approach aligns naturally with our theoretical analysis in Theorem~\ref{thm:HND_DA}. 

\myparagraph{Negative Log-Likelihood Loss}
Given labels $y_i\in\{1,\dots,C\}$, the negative log-likelihood loss is
\begin{equation}
\mathcal{L}_{\mathrm{NLL}}
= -\frac{1}{N}\sum_{i=1}^{N} \log\Big(\textit{softmax}(H_{\mathrm{final}}[i,:])_{y_i}\Big).
\end{equation}
It encourages high probability on the ground-truth class by penalizing low confidence on $y_i$.

\myblue{
\subsection{Extension Discussion}\label{discussion}
While AdaptCS is evaluated on undirected graphs with non-overlapping communities, the proposed model can be extended using standard techniques from prior work~\cite{huang2021sdgnn,smn_25}.

\myparagraph{Directed graphs}
Directed graphs can be handled mainly in the encoder by adopting standard directional message passing, e.g., separating {in/out} neighborhoods, followed by a fusion of these channels. 
This is consistent with SDGNN~\cite{huang2021sdgnn}, which is guided by status/balance theory and learns representations by jointly modeling link direction, link sign, and signed directed triads (triangles) as auxiliary supervision. 

\myparagraph{Graphs with labeled edges}
Graphs with labeled (or typed) edges can be supported by incorporating edge semantics into the encoder, e.g., using relation-specific transformations or edge-aware attention to modulate message passing for different edge types. 
In this way, the learned representations reflect both structural connectivity and edge-level semantics, while the online search procedure remains unchanged.

\myparagraph{Overlapping communities}
Our framework can be extended to overlapping community search by adopting the Sparse Subspace Filter (SSF) paradigm in SMN~\cite{smn_25}. 
This subspace-based retrieval can be integrated into our offline--online pipeline with minimal changes, while preserving efficient online search.
}

\section{Efficient Optimization}
\label{sec:mem_opt}

Although distinctive-hop channels successfully disentangle multi-hop semantics, explicitly computing exact-hop adjacency matrices $\hat{A}^{(k)}$ introduces severe memory limitations, especially on large or dense graphs.
Unlike standard multi-hop methods such as SGC~\cite{sgc_2019} and MixHop~\cite{Mixhop}, which efficiently precompute high-order features using iterative operations like $A(AX)$ and avoid ever materializing high-order adjacency matrices, our approach requires explicit computation of terms such as $(AA - A)X$ to isolate purely two-hop neighbors.
However, forming $AA$ results in significant memory overhead even when $A$ is sparse.
To address this critical bottleneck, we introduce a memory-efficient optimization using low-rank singular value decomposition (SVD) to approximate distinctive-hop computations in a compressed latent subspace, eliminating the need to construct high-order adjacency matrices explicitly.

\subsection{Low-rank distinctive-hop computation via SVD}
\myblue{
To avoid explicit high-order adjacency matrices, we adopt a rank-$r$ truncated SVD approximation of the normalized one-hop adjacency operator. Specifically, we factorize the symmetrically normalized adjacency (low-pass channel) as
\(
\hat{A}\approx U_{r}\Sigma_{r}V_{r}^{\!\top},
\quad \text{where}\quad
U_{r},V_{r}\in\mathbb{R}^{n\times r},\;
\Sigma_{r}=\operatorname{diag}(s_{1},\dots,s_{r}).
\)
For any hop $k\ge1$, the approximated matrix is as
\begin{equation}
\label{eq:svd}
\hat{A}^{k}\approx U_{r}\Sigma_{r}^{k}V_{r}^{\!\top},\quad\text{with}\quad
\Sigma_{r}^{k}\coloneq\operatorname{diag}\!\bigl(s_{1}^{k},\dots,s_{r}^{k}\bigr).
\end{equation}
Therefore, the distinctive-hop adjacency becomes
\begin{equation}
\hat{A}^{(k)}
=\hat{A}^{k}-\hat{A}^{k-1}
\approx
U_{r}\bigl(\Sigma_{r}^{k}-\Sigma_{r}^{k-1}\bigr)V_{r}^{\!\top}.
\label{eq:deltaA}
\end{equation}
Using this representation, the low-pass feature at hop $k$ is computed in the compressed subspace as
\begin{equation}
X^{(k)}
=\hat{A}^{(k)}X
\approx
U_{r}\Delta\Sigma_{r}^{(k)}\bigl(V_{r}^{\!\top}X\bigr),
\label{eq:lowpass}
\end{equation}
where \( V_{r}^{\!\top}X \in \mathbb{R}^{r\times d} \) is computed once and shared across hops.

}

This SVD-based approach yields two major advantages:
First, all distinctive-hop computations are performed in the low-rank latent space, so there is no need to explicitly construct or store any $n \times n$ high-order adjacency matrices. This completely removes the memory bottleneck and enables our model to scale to graphs with hundreds of millions of edges using only a single GPU.
Second, the computational process is highly efficient: only a single SVD decomposition and one feature projection ($V_r^{\top}X$) are required. All subsequent hop-specific computations are reduced to fast diagonal matrix multiplications in the small $r$-dimensional space, which greatly accelerates the overall computation and makes the method suitable for large-scale and high-order graph analysis.

\subsection{Global weight renormalization}
To further stabilize propagation within this compressed pipeline, where local weight normalization is infeasible, we introduce a fast node-level renormalization that reweights hop-$k$ features based on aggregated connection strengths:
\begin{equation}
\mathbf{w}^{(k)} = \sigma\!\left(U_r\Delta\Sigma_r^{(k)}V_r^{\!\top}\mathbf{1}\right),\quad
X^{(k)} = \mathbf{w}^{(k)} \odot X^{(k)},
\end{equation}
where $\mathbf{1}$ is an all-ones vector, $\sigma$ denotes the sigmoid function, and $\odot$ represents element-wise scaling.
This global normalization maintains the memory advantage of the SVD formulation while adaptively amplifying nodes whose receptive fields expand across hops, effectively preventing gradient collapse.

\section{Methodology Phase \romannumeral 2: Online Searching} \label{sec:online}
In the online search phase, we introduce two methods for efficiently retrieving communities from precomputed embeddings. 
The first method leverages a Signed Community Search (SCS) algorithm that operates on a positive signed graph constructed from node embeddings, and incorporates a teleportation step to effectively handle sparse connections by allowing direct jumps to nodes highly similar to the query. 
Adaptive Community Score (ACS) selects candidate nodes based on their embedding similarity to the query and then ranks them using a homophily-adaptive scoring function. 
Both methods are designed to handle the challenges presented by heterophilic graphs, ensuring accurate and meaningful community retrieval.


\subsection{Online Search via SCS}
We first solve the absence of edge signs in heterophilic graphs by using the learned node embeddings from the AdaptCS encoder to infer the sign of each edge. 
For each edge $(u, v)$ in the original graph, we compute the cosine similarity between their normalized embeddings; if the similarity exceeds a threshold $\tau$, we treat $(u, v)$ as a positive edge.
This yields a pruned, semantically meaningful subgraph that avoids introducing spurious edges between unconnected nodes.
Community search then proceeds via a BFS exploration over the positive graph, starting from the query node. 
Due to the predominant number of negative edges and sparsity in heterophilic graphs, BFS may frequently reach a dead-end before the desired community size $\mathcal{K}$ is reached. 
To address this, we use a teleportation mechanism: whenever BFS is exhausted but $|\mathcal{C}_q| < \mathcal{K}$, where $|\mathcal{C}_q|$ is the size of the current community, we select the unvisited node most similar to the query node (according to the learned embedding similarity) and resume BFS from there. 
Algorithm pseudocode is disclosed in Appendix A online \cite{adaptcs}.



\subsection{Online Search via ACS}

The Adaptive Community Score (ACS) algorithm retrieves semantically relevant and structurally coherent communities by combining embedding similarity with a homophily-adaptive bonus or penalty score on direct connections.  
For each query node, a candidate set is first selected based on embedding similarity.  
ACS then assigns a score to each candidate, combining: (i) its cosine similarity to the query node, and (ii) an adaptive reward or penalty if it is directly connected to the query, where the magnitude and sign depend on the estimated global homophily ratio.  
The final community is obtained by selecting the $\mathcal{K}$ candidates with the highest total scores.


\myparagraph{Homophily-adaptive scoring}
Given a query node $q$, normalized embeddings $H \in \mathbb{R}^{n\times h}$, adjacency matrix $A$, target community size $\mathcal{K}$, similarity-weighting parameter $\tau \in [0,1]$, scalar hyperparameters $\lambda_\text{bonus}$, $\lambda_\text{penalty}$, and top factor $\alpha$, the ACS score for a candidate node $u$ is computed as:
\begin{equation}\label{eq:acs}
\mathrm{ACS}(u) = \tau S_{qu} + (1-\tau) \cdot A_{qu} \cdot w(u),
\end{equation}
where $S_{qu}$ is the cosine similarity between $q$ and $u$, $A_{qu} = 1$ if $u$ is adjacent to $q$ and $0$ otherwise, and $w(u)$ is defined as
\begin{equation}\label{eq:bonus}
w(u) =
\begin{cases}
h_{edge} \cdot \lambda_\text{bonus},  & \text{if } h_{edge} \geq 0.5 \\[4pt]
- (1-h_{edge}) \cdot \lambda_\text{penalty}, & \text{if } h_{edge} < 0.5
\end{cases}
\end{equation}
where $h_{edge}$ is the estimated global homophily ratio.
\myblue{When $\tau=1$, only the similarity term is used and topology is ignored.
}



\myparagraph{ACS search procedure}
ACS first estimates the global homophily ratio and selects a candidate set by semantic similarity to the query node. For each candidate, it computes a homophily-adaptive score that combines similarity and, if directly connected to the query, a topology-aware bonus or penalty. The $\mathcal{K}$ highest-scoring nodes (plus the query itself) form the returned community. 
This method is robust across varying homophily levels and avoids the pitfalls of methods that rely solely on topology or similarity.
Algorithm pseudocode is disclosed in Appendix A online \cite{adaptcs}.

\section{Theoretical Analysis}\label{train & query}

\subsection{Distance Awareness Suppresses Similarity Flip}\label{sec:proof_HDD}

\myblue{

\begin{theorem}[Distance-aware hop separation suppresses similarity flip]
\label{thm:HND_DA}
Recall the HND gap $\Phi_u(\cdot)$ in Definition~\ref{def:HND}.
Let $z_u\in\{1,\dots,c\}$ be the class label of node $u$, and let $\{p_j\}_{j=1}^c\subset\mathbb{R}^d$ be a prototype bank with $\|p_j\|\le P$ for all $j$.
Given a root representation $r_u\in\mathbb{R}^d$ and exact-hop representations $h_u^{(1)},h_u^{(2)}\in\mathbb{R}^d$,
our fusion first forms class weights $\alpha^{class}_{u,j}$, then computes hop-attention weights $\alpha^{hop}_{u,k}$.
Assume $\|h_u^{(1)}\|,\|h_u^{(2)}\|\le B$ for all $u$.
If there exists an {informative hop} (here $k=2$) such that $\Phi_u(h^{(2)})\ge\delta$ for all $u$,
and the above bank-conditioned hop attention selects this hop for each node $u$ in the large-temperature regime,
then for sufficiently large $(\beta_c,\beta_h)$ we have $\Phi_u(h)>0$ for all $u$, hence $\mathrm{HND}(G)=1$.
Moreover, a sufficient condition for hop-$2$ selection is the existence of margins $\gamma,\eta>0$ such that
$\langle r_u,p_{z_u}\rangle\ge\max_{j\neq z_u}\langle r_u,p_j\rangle+\gamma$ and
$\langle h_u^{(2)},p_{z_u}\rangle\ge\langle h_u^{(1)},p_{z_u}\rangle+\eta$ for all $u$.
\end{theorem}

\myparagraph{Remark (Effect of feature noise)}
When the input features are noisy, e.g., $x_u = x_u^\star + \xi_u$ with $\|\xi_u\|\le \epsilon$.
Such bounded noise perturbs inner products by at most $O(\epsilon)$, and therefore reduces the margins
(e.g., $\gamma,\eta$) and the informative-hop gap (e.g., $\delta$) by a bounded amount.
As long as the perturbed margins and gap remain positive, the conclusion of Theorem~\ref{thm:HND_DA} still holds. 

\begin{proof}
We verify hop-$2$ selection under the stated sufficient margin conditions and then conclude HND.

\textbf{Step 1 (root margin $\Rightarrow w_u\approx p_{z_u}$).}
Let $z_u$ be the true class of $u$ and recall
$\alpha^{class}_{u,j}\propto \exp(\beta_c\langle r_u,p_j\rangle)$.
If $\langle r_u,p_{z_u}\rangle\ge\max_{j\neq z_u}\langle r_u,p_j\rangle+\gamma$, then for any $j\neq z_u$,
\[
\frac{\alpha^{class}_{u,j}}{\alpha^{class}_{u,z_u}}
=\exp\!\big(\beta_c(\langle r_u,p_j\rangle-\langle r_u,p_{z_u}\rangle)\big)
\le e^{-\beta_c\gamma}.
\]
Thus $\sum_{j\neq z_u}\alpha^{class}_{u,j}\le (c-1)e^{-\beta_c\gamma}$ and
\[
\|w_u-p_{z_u}\|\le 2P\sum_{j\neq z_u}\alpha^{class}_{u,j}\le 2P(c-1)e^{-\beta_c\gamma}.
\]
so $w_u = p_{z_u} + O\!\big(P(c-1)e^{-\beta_c\gamma}\big)$.

\textbf{Step 2 (hop margin $\Rightarrow \alpha^{hop}_{u,2}\approx 1$).}
For $k\in\{1,2\}$, since $\|h_u^{(k)}\|\le B$ and $\|w_u-p_{z_u}\|\le 2P(c-1)e^{-\beta_c\gamma}$,
\[
\big|\langle h_u^{(k)},w_u\rangle-\langle h_u^{(k)},p_{z_u}\rangle\big|
\le 2BP(c-1)e^{-\beta_c\gamma}.
\]
If additionally $\langle h_u^{(2)},p_{z_u}\rangle\ge\langle h_u^{(1)},p_{z_u}\rangle+\eta$, then
\[
\langle h_u^{(2)},w_u\rangle-\langle h_u^{(1)},w_u\rangle
\ge \eta - 4BP(c-1)e^{-\beta_c\gamma}.
\]
For sufficiently large $\beta_c$ such that $4BP(c-1)e^{-\beta_c\gamma}\le \eta/2$, we have
$\langle h_u^{(2)},w_u\rangle \ge \langle h_u^{(1)},w_u\rangle + \eta/2$.
Since $\alpha^{hop}_{u,k}\propto\exp(\beta_h\langle h_u^{(k)},w_u\rangle)$, this implies
\[
\Delta_u:=\langle h_u^{(2)},w_u\rangle-\langle h_u^{(1)},w_u\rangle \ge \eta/2
\]
\[\Rightarrow
\alpha^{hop}_{u,2}=\frac{1}{1+e^{-\beta_h\Delta_u}}\ge \frac{1}{1+e^{-\beta_h\eta/2}}.
\]

Therefore, using $h_u=\sum_{k\in\{1,2\}}\alpha^{hop}_{u,k}h_u^{(k)}$ and $\|h_u^{(k)}\|\le B$,
\[
\|h_u-h_u^{(2)}\|\le 2B(1-\alpha^{hop}_{u,2})\le 2B e^{-\beta_h\eta/2}.
\]

\textbf{Step 3 (stability of the HND gap).}
For inner-product similarity (or $\epsilon$-stabilized cosine), the similarity $S(u,v)$ and hence the induced gap $\Phi_u(\cdot)$
vary continuously on the bounded set $\{\|h\|\le B\}$.
Thus the perturbation $\|h_u-h_u^{(2)}\|\le 2B e^{-\beta_h\eta/2}$ implies
\[
\Phi_u(h)\ge \Phi_u(h^{(2)}) - O\!\big(e^{-\beta_h\eta/2}\big).
\]
Using the informative-hop condition $\Phi_u(h^{(2)})\ge \delta$, we obtain
\[
\Phi_u(h)\ge \delta - O\!\big(e^{-\beta_h\eta/2}\big).
\]
Having $\beta_h$ sufficiently makes the error smaller than $\delta$,
so $\Phi_u(h)>0$ for all $u$, which satisfy the HND condition.
\end{proof}

}

\subsection{Time Complexity Analysis}



\myblue{
\myparagraph{Query efficiency comparison}
SCS costs $O(n+m)$ per query, while ACS costs $O(nh + M\log M)$, dominated by the similarity scan.
Full analysis is in Appendix A online \cite{adaptcs}.

\noindent {ICSGNN/QDGNN/COCLEP} are {query-driven}: each query requires a query-conditioned forward propagation over the graph. Notably, the original ICSGNN implementation is {fully online} (i.e., it trains for each query without a separate offline stage); for fair comparison, we restructured it into an offline-training plus online-search pipeline similar to QDGNN. However, in these methods, each query still requires $k$-layer inference, giving a shared per-query complexity
\(
O\!\big(k(mh+nh^2)\big).
\)

\noindent {CommunityDF} iteratively refine on a query-induced subgraph $G_s=(V_s,E_s)$: it first computes a distribution (e.g., $O(|E_s|+|V_s|\log|V_s|)$ via PPR or $O(L(|E_s|h+|V_s|h^2))$ via GCN scoring) and then runs $T_c$ diffusion steps. Thus, its per-query cost scales with $T_c$ and can be summarized as
\(
O\!\big(|E_s|+|V_s|\log|V_s|\big)\;+\;O\!\big(T_c\,L(|E_s|h+|V_s|h^2)\big),
\)
where $T_c\ll T$ is the reduced diffusion steps in their optimization.

\noindent Overall, AdaptCS/ACS is {query-decoupled}: it avoids per-query graph-wide $k$-layer inference and only performs a similarity scan plus candidate ranking.
In contrast, query-driven methods incur $O(k(mh+nh^2))$ per query due to graph propagation, and CommunityDF further adds iterative refinement whose cost grows with $T_c$ on the query-induced subgraph.
}

\begin{table*}[t]
    \caption{Effectiveness evaluation of different datasets} \label{tab:data}
    \centering
    \resizebox{\textwidth}{!}{
    \begin{tabular}{c|c||cccccccc|cccccccc|c}
        \toprule
         \multicolumn{2}{c||}{\textbf{Baselines}} & 
        \multicolumn{8}{c|}{\textbf{Homophilic graphs}} & 
        \multicolumn{8}{c|}{\textbf{Heterophilic graphs}} & 
        \multicolumn{1}{c}{\textbf{Average}} \\
            \midrule
        Models & Extensions & Cora & CiteSeer & Photo & Computers & DBLP & CS & PubMed & Reddit & Cornell & Texas & Wisconsin & Chamel & Squirrel & Film & Roman & Flickr & Ave +/-  \\ 
            \midrule
            \midrule
        k-core & Vanilla & 0.3682 & 0.3567 & 0.5781 & 0.5069 & 0.689 & 0.2922 & 0.6016 & 0.2967 & 0.6404 & 0.6854 & 0.5895 & 0.4253 & 0.4745 & 0.3758 & 0.2037 & 0.5032 & -40.91\%  \\ 
        CTC & Vanilla & 0.3574 & 0.3303 & 0.5253 & 0.5039 & 0.6705 & 0.2942 & 0.5933 & 0.1650 & 0.7007 & 0.6611 & 0.6190 & 0.4121 & 0.6590 & 0.3852 & 0.2328 & 0.5512 & -40.33\%  \\ 
        k-clique & Vanilla & 0.3716 & 0.3183 & 0.4712 & 0.4781 & 0.6446 & 0.2640 & 0.5743 & - & 0.6806 & 0.7419 & 0.6334 & 0.4898 & 0.6098 & 0.3813 & 0.2838 & - & -38.20\%  \\ 
            \midrule
        LP & Vanilla & 0.8147 & 0.7266 & 0.8444 & 0.7136 & 0.8603 & 0.8682 & 0.7702 & - & 0.7410 & 0.7571 & 0.6814 & 0.4618 & 0.4895 & 0.3813 & 0.2552 & - & -16.64\%  \\ 
        LP & Attributed & 0.8586 & 0.8592 & \underline{0.8450} & 0.7210 & 0.8894 & 0.8222 & 0.7824 & - & 0.8056 & 0.8139 & 0.8614 & 0.4594 & 0.4484 & 0.3808 & 0.3936 & - & -11.51\%  \\ 
            \midrule
        ICSGNN & Tanh & 0.7874 & 0.7764 & 0.7341 & 0.7771 & 0.8604 & 0.8521 & 0.8048 & - & 0.5977 & 0.6524 & 0.6111 & 0.4327 & 0.4414 & 0.3711 & 0.2053 & 0.4137 & -22.59\%  \\ 
        \myblue{ICSGNN} & \myblue{Mixhop} & \myblue{0.7312} & \myblue{0.5858} & \myblue{0.8180} & \myblue{0.7306} &
        \myblue{0.7811} & \myblue{0.6740} & \myblue{0.7907} & \myblue{-} & \myblue{0.5314} & \myblue{0.6478} &
        \myblue{0.6191} & \myblue{0.4115} & \myblue{0.5020} & \myblue{0.3832} & \myblue{0.2410} & \myblue{0.3805} & \myblue{-26.66\%} \\
        \myblue{ICSGNN} & \myblue{Geom} & \myblue{0.7414} & \myblue{0.7222} & \myblue{0.7222} & \myblue{0.7851} &
        \myblue{0.8106} & \myblue{0.8359} & \myblue{0.8520} & \myblue{-} & \myblue{0.5703} & \myblue{0.6794} &
        \myblue{0.6003} & \myblue{0.5109} & \myblue{0.4595} & \myblue{0.3780} & \myblue{0.4088} & \myblue{-} & \myblue{-19.21\%} \\
        ICSGNN & ACM & 0.8500 & 0.7086 & 0.7397 & 0.7599 & 0.8525 & 0.8811 & 0.8420 & - & 0.5963 & 0.7145 & 0.6098 & 0.4327 & 0.4414 & 0.3446 & 0.3339 & 0.4197 & -20.85\%  \\ 
        ICSGNN & ALT & 0.8133 & 0.7711 & 0.3243 & 0.3525 & 0.8828 & - & 0.8425 & - & 0.6391 & 0.7696 & 0.7128 & 0.4327 & 0.4414 & 0.3693 & 0.3516 & - & -26.16\%  \\ 
            \midrule
        QDGNN & Tanh & 0.8382 & 0.8078 & 0.7921 & 0.8319 & 0.8419 & 0.8871 & 0.8458 & - & 0.7140 & 0.6493 & 0.5990 & 0.3757 & 0.4668 & 0.3813 & 0.1899 & - & -17.92\%  \\ 
        \myblue{QDGNN} & \myblue{Mixhop} & \myblue{0.7838} & \myblue{0.7445} & \myblue{0.7216} & \myblue{0.8018} &
        \myblue{0.8595} & \myblue{0.7358} & \myblue{0.8425} & \myblue{-} & \myblue{0.7654} & \myblue{0.8395} &
        \myblue{0.6157} & \myblue{0.4046} & \myblue{0.4595} & \myblue{0.3698} & \myblue{0.3314} & \myblue{-} & \myblue{-17.44\%} \\
        \myblue{QDGNN} & \myblue{Geom} & \myblue{0.7529} & \myblue{0.6936} & \myblue{0.7326} & \myblue{0.7993} &
        \myblue{0.7682} & \myblue{0.8278} & \myblue{0.8753} & \myblue{-} & \myblue{0.7742} & \myblue{0.7794} &
        \myblue{0.6223} & \myblue{0.4771} & \myblue{0.4668} & \myblue{0.3846} & \myblue{0.3681} & \myblue{-} & \myblue{-16.13\%} \\
        QDGNN & ACM & 0.8721 & 0.8615 & 0.6669 & 0.7375 & 0.8045 & - & 0.7848 & - & 0.6734 & 0.8560 & 0.7617 & 0.3757 & 0.4668 & 0.3319 & 0.3384 & - & -18.22\%  \\ 
        QDGNN & ALT & 0.8534 & 0.6598 & 0.5829 & - & 0.7642 & - & 0.8231 & - & 0.5471 & 0.6593 & 0.5302 & 0.3757 & 0.4668 & 0.3389 & 0.3498 & - & -27.81\%  \\ 
            \midrule
        COCLEP & Tanh & 0.5167 & 0.7553 & 0.8042 & - & - & - & 0.8510 & - & 0.5405 & 0.6223 & 0.5629 & 0.3757 & 0.4668 & 0.3404 & 0.2210 & - & -31.38\%  \\ 
        \myblue{COCLEP} & \myblue{Mixhop} & \myblue{0.5482} & \myblue{0.6702} & \myblue{0.6494} & \myblue{-} &
        \myblue{-} & \myblue{-} & \myblue{0.8478} & \myblue{-} & \myblue{0.5866} & \myblue{0.6501} &
        \myblue{0.5727} & \myblue{0.3346} & \myblue{0.4882} & \myblue{0.3716} & \myblue{0.2651} & \myblue{-} & \myblue{-32.20\%} \\
        \myblue{COCLEP} & \myblue{Geom} & \myblue{0.4460} & \myblue{0.6399} & \myblue{0.6781} & \myblue{-} &
        \myblue{-} & \myblue{-} & \myblue{0.8783} & \myblue{-} & \myblue{0.5670} & \myblue{0.6712} &
        \myblue{0.5311} & \myblue{0.3253} & \myblue{0.4268} & \myblue{0.3705} & \myblue{0.2554} & \myblue{-} & \myblue{-34.41\%} \\
        COCLEP & ACM & 0.3676 & 0.3615 & 0.3173 & - & - & - & 0.8665 & - & 0.5427 & 0.7314 & 0.5390 & 0.3757 & 0.4668 & 0.3313 & 0.2731 & - & -41.40\%  \\ 
        COCLEP & ALT & 0.4627 & 0.4363 & 0.6108 & - & - & - & 0.7024 & - & 0.5287 & 0.6011 & 0.5227 & 0.3757 & 0.4668 & 0.3395 & 0.4368 & - & -37.88\%  \\ 
            \midrule
        ComDF & Tanh & 0.8487 & 0.7315 & 0.6246 & 0.6972 & 0.8730 & 0.7817 & 0.7660 & - & 0.6642 & 0.6996 & 0.6282 & 0.3700 & 0.3624 & 0.3678 & 0.2250 & - & -23.09\%  \\ 
        \myblue{ComDF} & \myblue{Mixhop} & \myblue{0.7256} & \myblue{0.6650} & \myblue{0.7822} & \myblue{0.6722} &
        \myblue{0.8525} & \myblue{0.7807} & \myblue{0.7217} & \myblue{-} & \myblue{0.6556} & \myblue{0.6334} &
        \myblue{0.6431} & \myblue{0.2967} & \myblue{0.3376} & \myblue{0.3742} & \myblue{0.2661} & \myblue{-} & \myblue{-25.17\%} \\
        \myblue{ComDF} & \myblue{Geom} & \myblue{0.7599} & \myblue{0.7043} & \myblue{0.7772} & \myblue{0.6204} &
        \myblue{0.8273} & \myblue{0.8233} & \myblue{0.7847} & \myblue{-} & \myblue{0.6686} & \myblue{0.6183} &
        \myblue{0.6069} & \myblue{0.4840} & \myblue{0.3666} & \myblue{0.3792} & \myblue{0.2991} & \myblue{-} & \myblue{-22.38\%} \\
        ComDF & ACM & 0.7452 & 0.6955 & 0.6905 & 0.6146 & 0.7716 & 0.6998 & 0.7717 & - & 0.6642 & 0.7042 & 0.7059 & 0.3689 & 0.3355 & 0.4276 & 0.3950 & - & -23.53\%  \\ 
        ComDF & ALT & 0.8377 & 0.7453 & 0.6905 & 0.6972 & 0.8309 & 0.8060 & 0.7833 & - & 0.6642 & 0.6960 & 0.6282 & 0.3689 & 0.3373 & 0.3781 & 0.2726 & - & -22.24\%  \\ 
            \midrule
        AdaptCS-I & ACS & \textbf{0.9089} & \underline{0.9219} & \textbf{0.9046} & \textbf{0.8522} & \textbf{0.9145} & \textbf{0.9370} & \textbf{0.9544} & - & \underline{0.8897} & \underline{0.8885} & \underline{0.8948} & \textbf{0.5441} & \textbf{0.7846} & \textbf{0.4862} & \textbf{0.7238} & - & 0.00\%  \\ 
        AdaptCS-II & SCS & \underline{0.9095} & \textbf{0.9236} & 0.8260 & \underline{0.8369} & \underline{0.9044} & \underline{0.9345} & \underline{0.9079} & \underline{0.5577} & \textbf{0.8961} & \textbf{0.8969} & \textbf{0.9019} & \underline{0.5224} & \underline{0.7813} & \underline{0.4821} & \underline{0.7222} & \underline{0.5584} & -2.16\%  \\ 
        AdaptCS-II & ACS & \textbf{0.9089} & \underline{0.9219} & \textbf{0.9046} & \textbf{0.8522} & \textbf{0.9145} & \textbf{0.9370} & \textbf{0.9544} & \textbf{0.6642} & \underline{0.8897} & \underline{0.8885} & \underline{0.8948} & \textbf{0.5441} & \textbf{0.7846} & \textbf{0.4862} & \textbf{0.7238} & \textbf{0.5696} & 0.00\%  \\ 
            \bottomrule
    \end{tabular}
    }
\vspace{-7mm}
\end{table*}

\section{Experiments} \label{Experiments}
\subsection{Experimental Setup}

\myparagraph{Datasets}
We evaluate AdaptCS on 16 real-world graphs with varying levels of homophily.
Eight datasets exhibit high homophily, including Cora, CiteSeer, PubMed~\cite{cora_yang2016revisiting}, Amazon-Computers, Amazon-Photo, Coauthor-CS~\cite{shchur2018pitfalls}, DBLP~\cite{tang2008arnetminer}, and Reddit~\cite{reddit}.
The others show heterophilic patterns, including Cornell, Wisconsin, Texas, Film~\cite{geom-gcn}, Chameleon (corrected), Squirrel (corrected), Roman~\cite{platonovcritical}, and Flickr~\cite{factorLP_2020_SIGMOD}.
Statistics are summarized in Appendix B online \cite{adaptcs}.

\myparagraph{Baseline models}
We compare AdaptCS against three categories of baselines:
(i) {Algorithm-based methods} including $k$-core~\cite{k-core}, CTC~\cite{k-truss2015}, and $k$-clique~\cite{k-clique};
(ii) {H-based Label propagation (LP)} methods, evaluated in both vanilla and attributed settings, serving as non-parametric embedding-free baselines;
and (iii) {ML-based community search (ML-CS)} models, including ICSGNN~\cite{ics-gnn}, QDGNN~\cite{qdgnn}, COCLEP~\cite{coclep}, and ComDF~\cite{chen2025communitydf}.
\myblue{Since these ML-CS models were developed under the homophily assumption, we evaluate each baseline under a consistent set of heterophily-oriented extensions to ensure fair comparison across both homophilic and heterophilic graphs, including {Tanh}~\cite{FAGCN21}, {MixHop}~\cite{Mixhop}, {Geom-GCN}~\cite{geom-gcn}, {ACM}~\cite{ACM_2022}, and {ALT}~\cite{ALT_KDD_2023}.}
\myblue{We adopt the original hyperparameter settings reported in the respective baseline papers.}

\myparagraph{Evaluation metrics}
The evaluation of identified communities is conducted through F1-score~\cite{ics-gnn,qdgnn}.
The F1-score balances precision and recall, offering a measure of how well the identified community matches the ground truth. 
The true data is established as the target community label, with the labels of the identified nodes serving as the predicted data. 
To evaluate the efficiency, the model training time and online querying time are recorded across different models. 
All results are averaged across 50 randomly selected queries. 

\myparagraph{Implementation details}
\myblue{
The ground-truth community of a query is defined as the set of nodes sharing the same label as the query node.
Following the standard supervised settings~\cite{smn_25,qdgnn}, we split labeled nodes into train/val/test as 60\%/20\%/20\%.
The training set label is only used to compute the loss and update model parameters. 
During validation, the parameters are frozen, and the test set remains untouched to prevent information leakage.
All query nodes are randomly sampled from the test set only. 
}
We run AdaptCS for 100 epochs with early stopping.
AdaptCS uses a 5-hop receptive field and 512 hidden units by default. 
We use a 3-hop receptive field and 128 hidden units for large datasets to accommodate their scale.
\myblue{The attention-bank fusion is used for hop channel mixing.}
The learning rate is set as 0.01.
The community size $\mathcal{K}$ is dataset-dependent and may vary according to user needs. We set $\mathcal{K} = 30$ for small datasets, $150$ for medium-sized datasets, and $1000$ for large datasets.
The experiments are run on a machine with Intel Xeon  6248R CPU, Nvidia A5000 GPU, and 512GB memory.
The code is available at Github \footnote{https://github.com/SimonQS/AdaptCS}.

\subsection{Effectiveness of AdaptCS}
\autoref{tab:data} benchmarks our approach on 16 graphs, where {AdaptCS-I} is the vanilla model that performs distance- and frequency-aware aggregation on the full adjacency;  
{AdaptCS-II} adds the proposed {low-rank SVD} compression, which keeps hop-distinct features in the latent space and thus avoids memory overflow.  
For the online search phase, we compare two method: the Signed Community Search ({SCS}) and Adaptive Community Score ({ACS}).
“\texttt{-}” indicates that the method exceeds the GPU memory limit or fails to finish within the 12-hour wall-clock time.
The rightmost column, {Ave +/-}, reports the average F1-score difference of each method relative to the best AdaptCS-II (ACS) result on average across all datasets.

\myparagraph{Observation 1 – Algorithm-based methods are ineffective}
Structure-only algorithms ($k$-clique, CTC, $k$-core) achieve relatively lower F1 due to lacking explicit edge semantics.
Interestingly, their performance on heterophilic graphs is slightly better compared to homophilic graphs. 
A possible explanation is that balance theory implies that dense regions tend to minimize triadic tension, so a structural filter may still include useful members alongside noise.

\myparagraph{Observation 2 – Label propagation remains competitive}
Despite its simplicity, the label propagation (LP) baseline performs competitively on most homophilic datasets, often achieving scores comparable to or even exceeding those of shallow GNNs.
This suggests that diffusion-based similarity effectively captures meaningful community signals when edge connections align with class consistency.
However, its performance degrades substantially on heterophilic graphs, where local patterns provide limited insight, highlighting the inherent limitation of global smoothing in non-homophilic settings.

\myblue{\myparagraph{Observation 3 – Homophily-oriented GNN backbones remain limited}
Among the four ML-CS backbones, QDGNN achieves the best overall average (with its strongest variant under Geom-GCN), followed by ICSGNN and ComDF, while COCLEP consistently underperforms.
With the expanded set of five heterophily-oriented extensions/backbones (Tanh, MixHop, Geom-GCN, ACM, ALT), we observe that Geom-GCN and Tanh tend to yield the most consistent improvements for message-passing baselines, while ACM remains a competitive and stable alternative; ALT provides mixed gains and can even degrade performance for some backbones (with ALT also incurring higher overhead).
Nevertheless, even the best-configured baseline still falls short of AdaptCS.}

\myparagraph{Observation 4 – {AdaptCS} achieves the best and most stable performance}
Both {AdaptCS-I} and {AdaptCS-II} outperform all baselines across homophilic and heterophilic graphs.
Notably, their results are numerically identical under the same community search algorithm (ACS), demonstrating that the low-rank SVD optimization in {AdaptCS-II} effectively approximates the dominant structural patterns without sacrificing accuracy.
This confirms that the compressed subspace preserves essential spectral information while significantly improving scalability, while completely eliminating memory overflow on large graphs such as Reddit.
Meanwhile, ACS consistently surpasses SCS with 2.16\% on average, highlighting the benefit of adaptively balancing learned embeddings and topological cues.
In summary, AdaptCS-II (ACS) achieves the best performance, scales to large and dense graphs, and delivers high-quality communities for both heterophilic and homophilic graphs.

\begin{figure*}[t]
    \centering
    \begin{subfigure}[b][4cm][c]{0.49\linewidth}
        \centering
        \includegraphics[width=\linewidth]{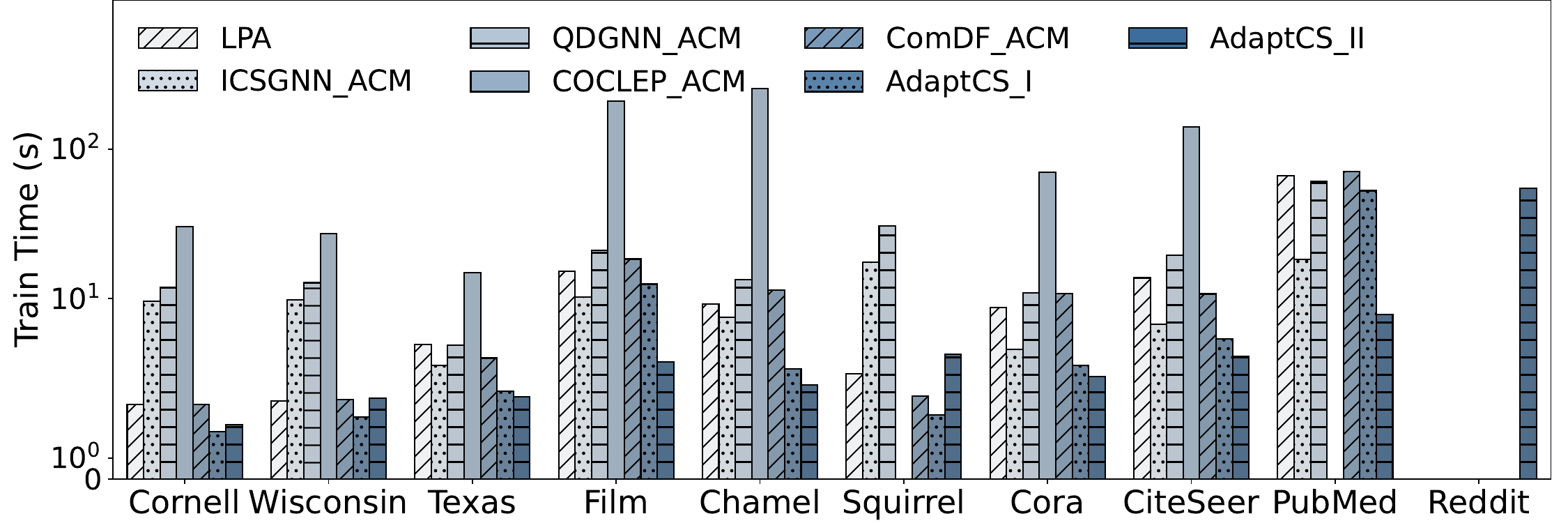}
        \caption{\myblue{Efficiency results of the training phase}}
        \label{fig:efficiency_1}
    \end{subfigure}
    \begin{subfigure}[b][4cm][c]{0.49\linewidth}
        \centering
        \includegraphics[width=\linewidth]{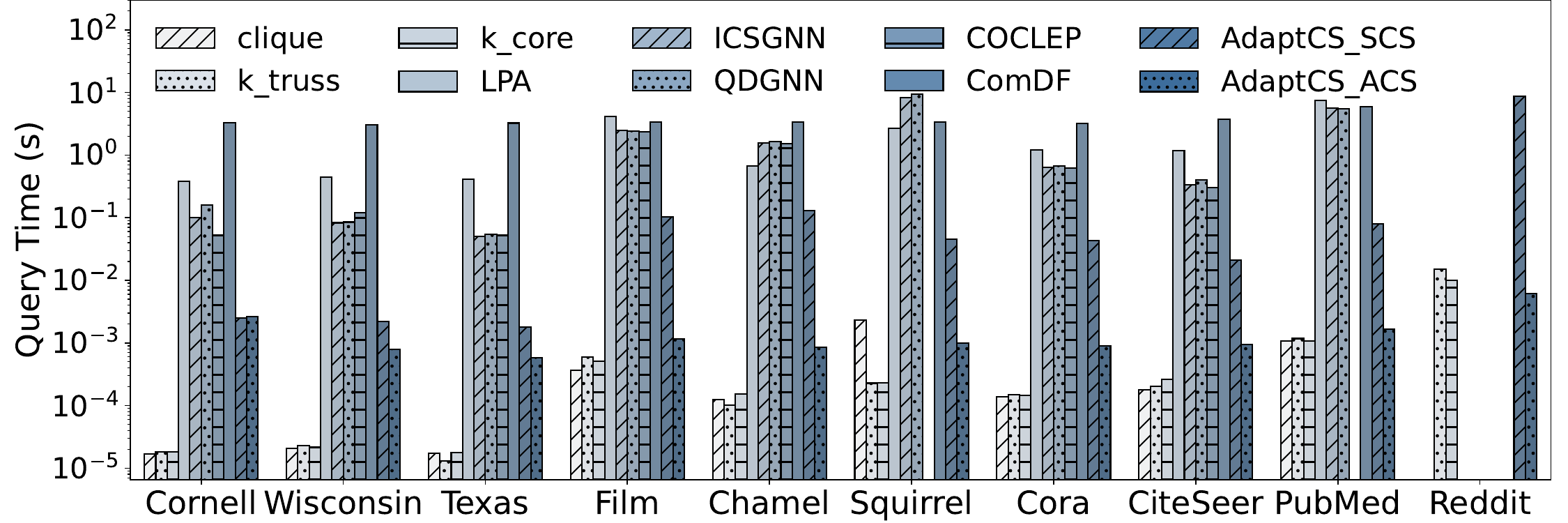}
        \caption{\myblue{Efficiency results of the query phase (per query)}}
        \label{fig:efficiency_2}
    \end{subfigure}
        \vspace{-5mm}
    \caption{\myblue{Efficiency evaluation of different datasets (in seconds)}}
        \vspace{-5mm}
\label{fig:efficiency}        
\end{figure*}

\begin{figure}[t]
  \centering
  \includegraphics[width=0.5\textwidth]{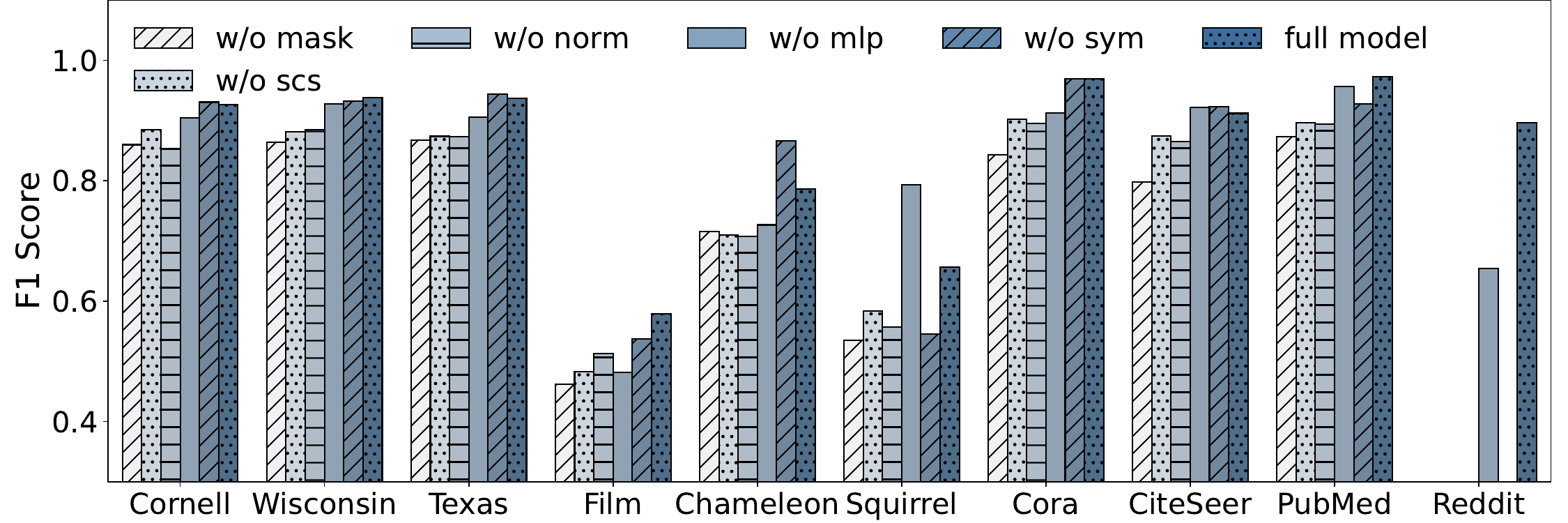}
        \vspace{-6mm}
  \caption{Ablation study}
        \vspace{-2mm}
  \label{fig:ablation}
\end{figure}

\begin{figure*}[t]
    \centering
    \begin{subfigure}[b][4cm][c]{0.32\linewidth}
        \centering
        \includegraphics[width=\linewidth]{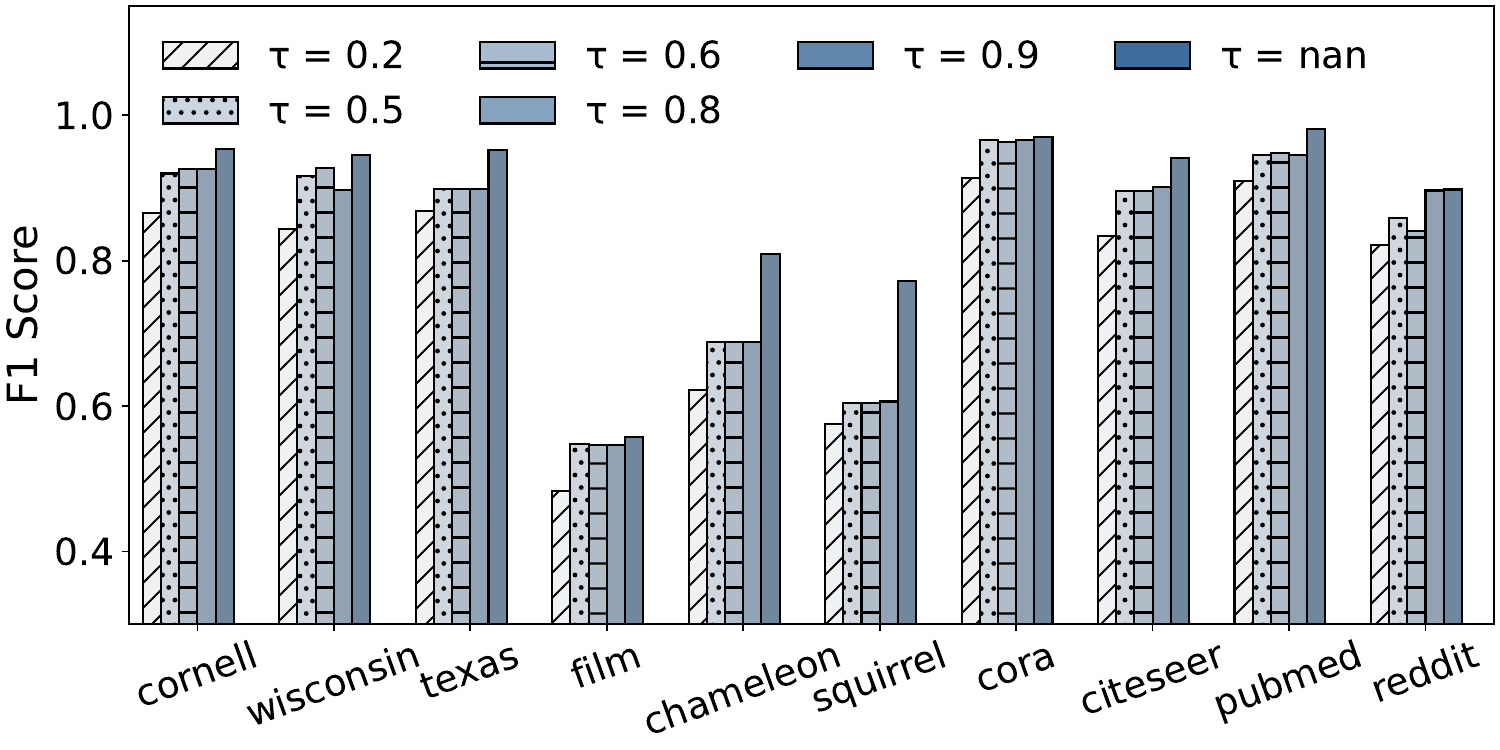}
        \vspace{-6mm}
        \caption{F1-Score by varying threshold $\tau$}
        \vspace{-20mm}
        \label{fig:chart1}
    \end{subfigure}
    \hfill
    \begin{subfigure}[b][4cm][c]{0.32\linewidth}
        \centering
        \includegraphics[width=\linewidth]{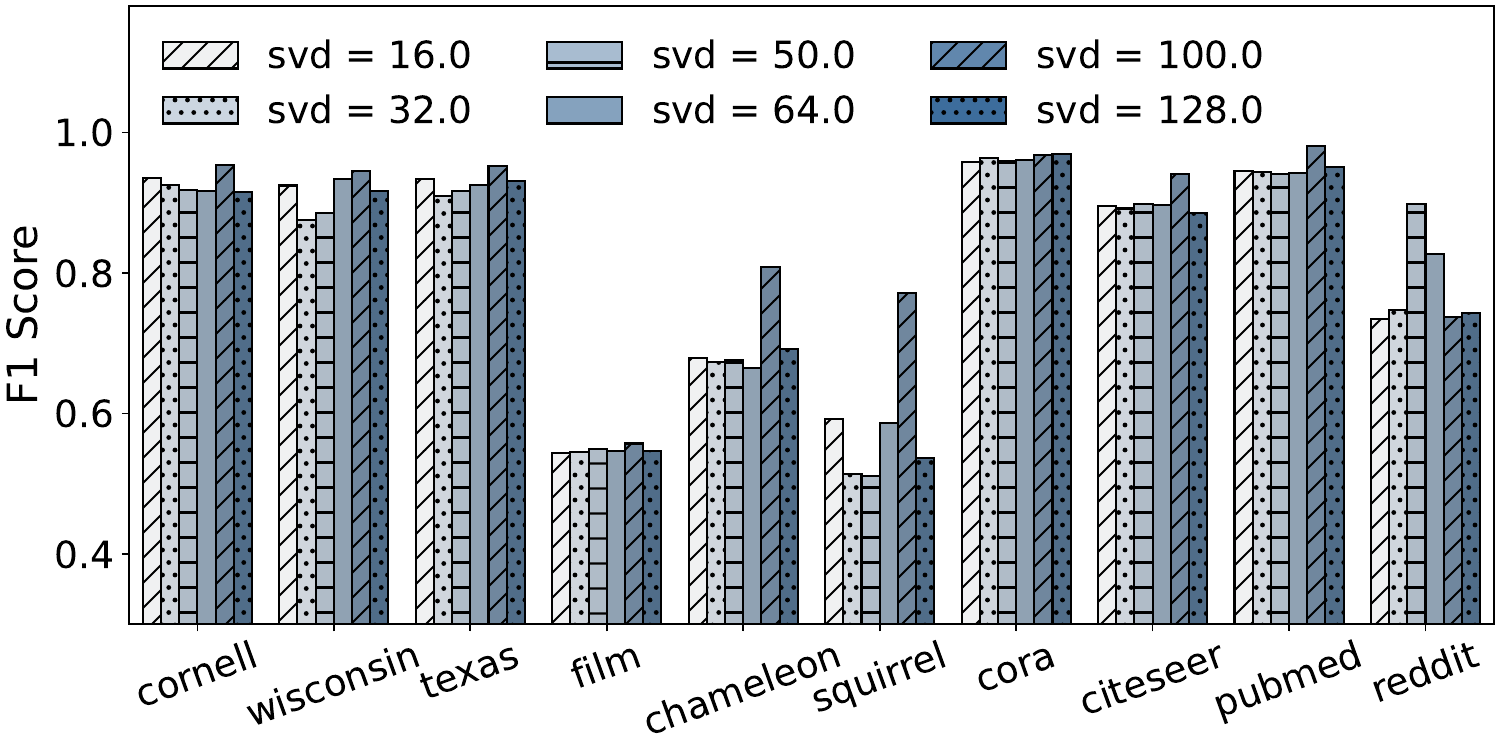}
        \vspace{-6mm}
        \caption{F1-Score by varying SVD rank}
        \vspace{-20mm}
        \label{fig:chart2}
    \end{subfigure}
    \hfill
    \begin{subfigure}[b][4cm][c]{0.32\linewidth}
        \centering
        \includegraphics[width=\linewidth]{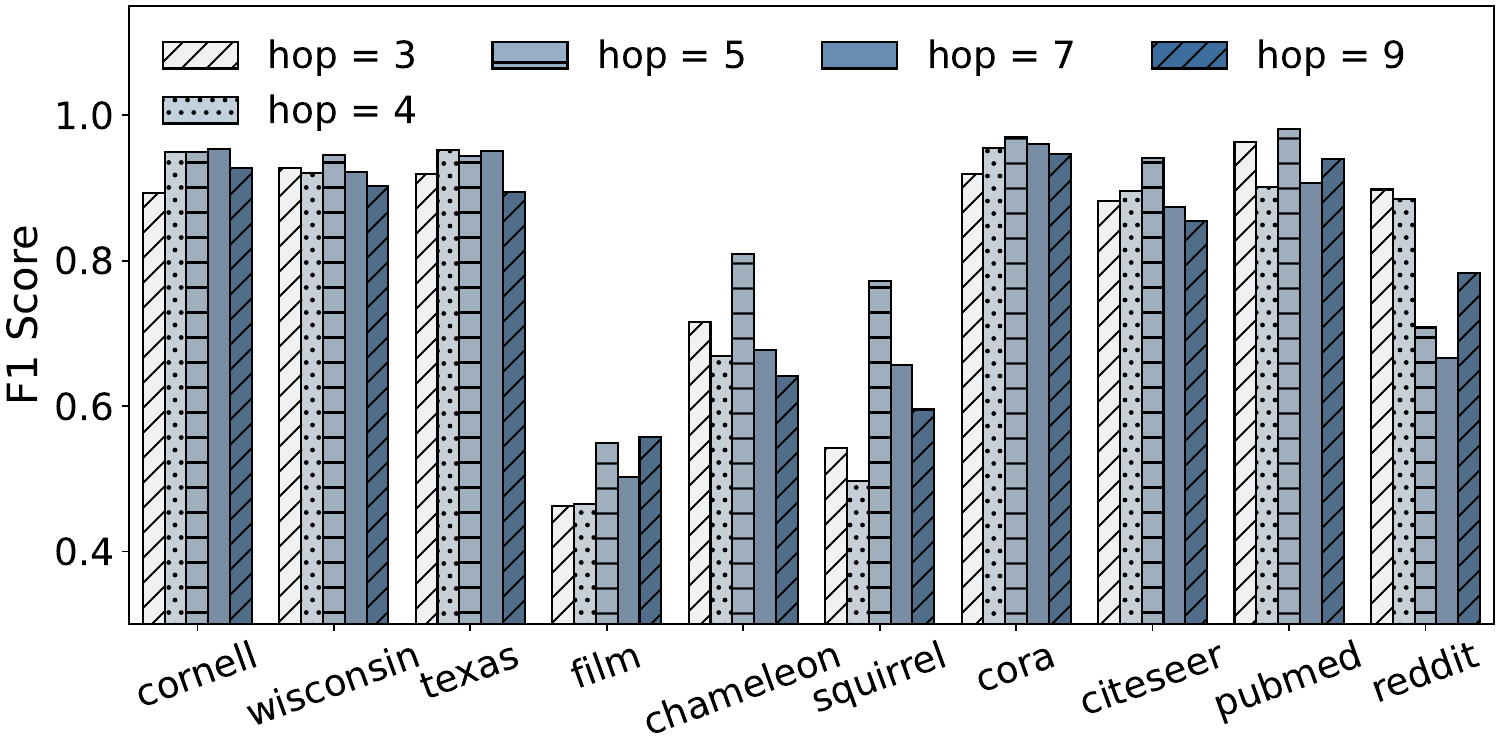}
        \vspace{-6mm}
        \caption{F1-Score by varying hop number}
        \vspace{-20mm}
        \label{fig:chart3}
    \end{subfigure}
    \centering
    \begin{subfigure}[b][4cm][c]{0.32\linewidth}
        \centering
        \includegraphics[width=\linewidth]{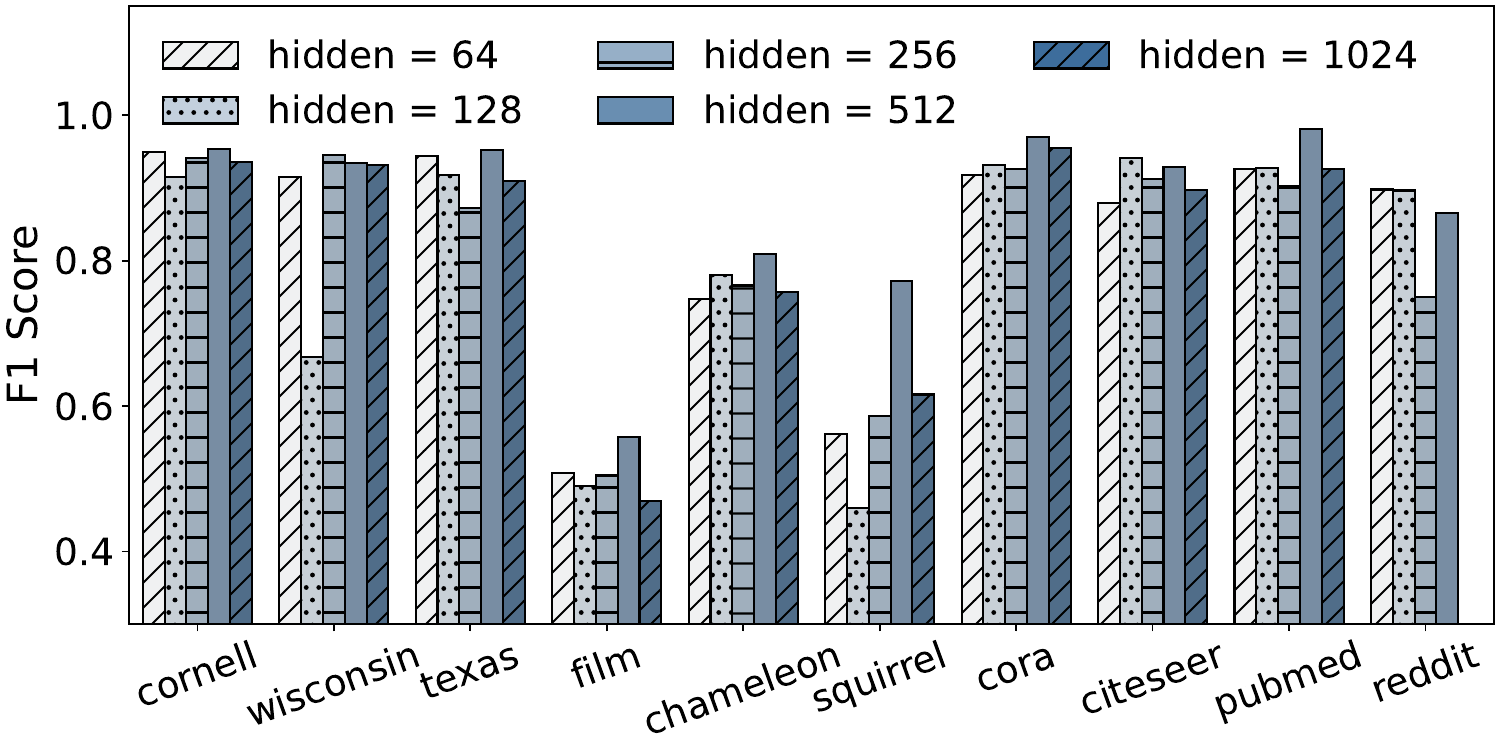}
        \vspace{-6mm}
        \caption{F1-Score by varying hidden state}
        \label{fig:chart4}
        \vspace{-6mm}
    \end{subfigure}
    \hfill
    \begin{subfigure}[b][4cm][c]{0.32\linewidth}
        \centering
        \includegraphics[width=\linewidth]{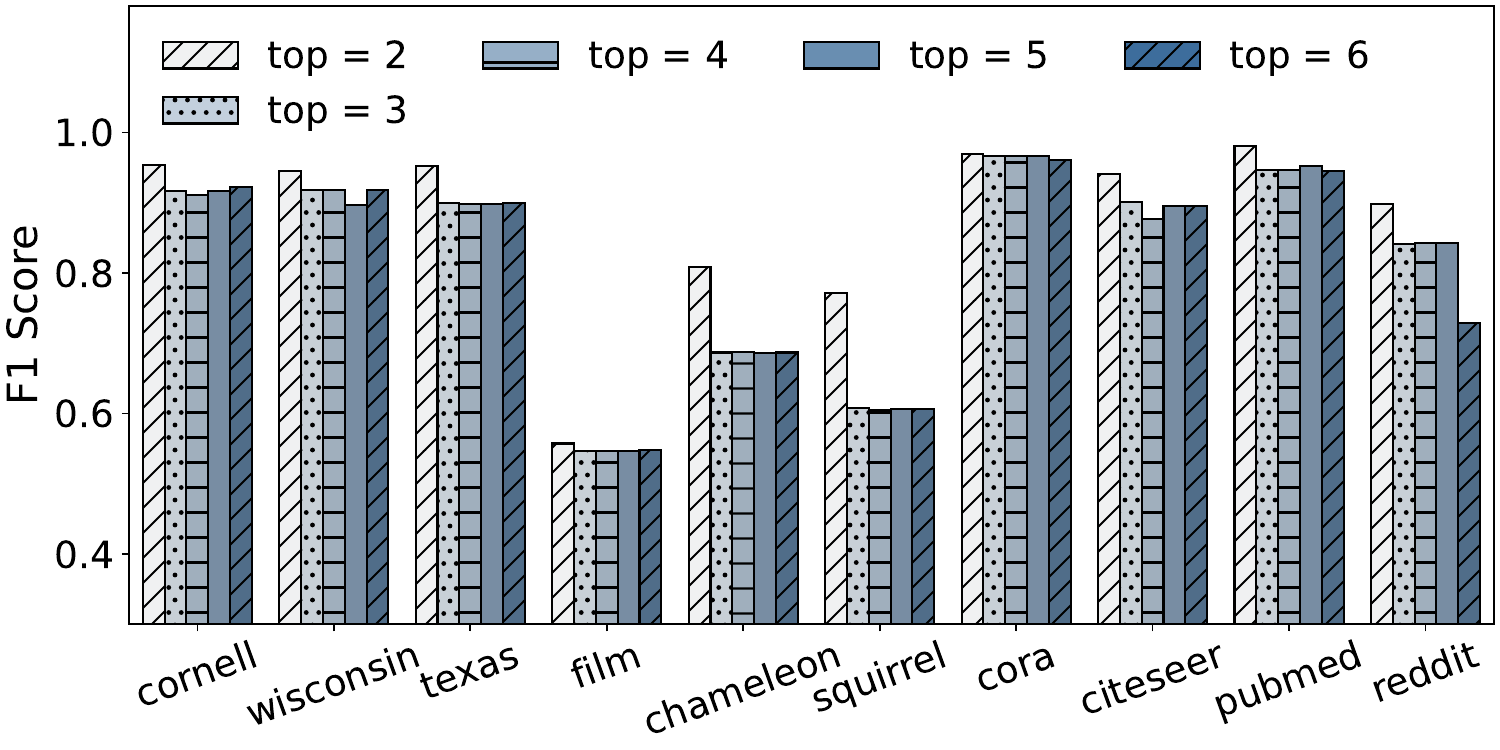}
        \vspace{-6mm}
        \caption{F1-Score by varying top factor}
        \label{fig:chart5}
        \vspace{-6mm}
    \end{subfigure}
    \hfill
    \begin{subfigure}[b][4cm][c]{0.32\linewidth}
        \centering
        \includegraphics[width=\linewidth]{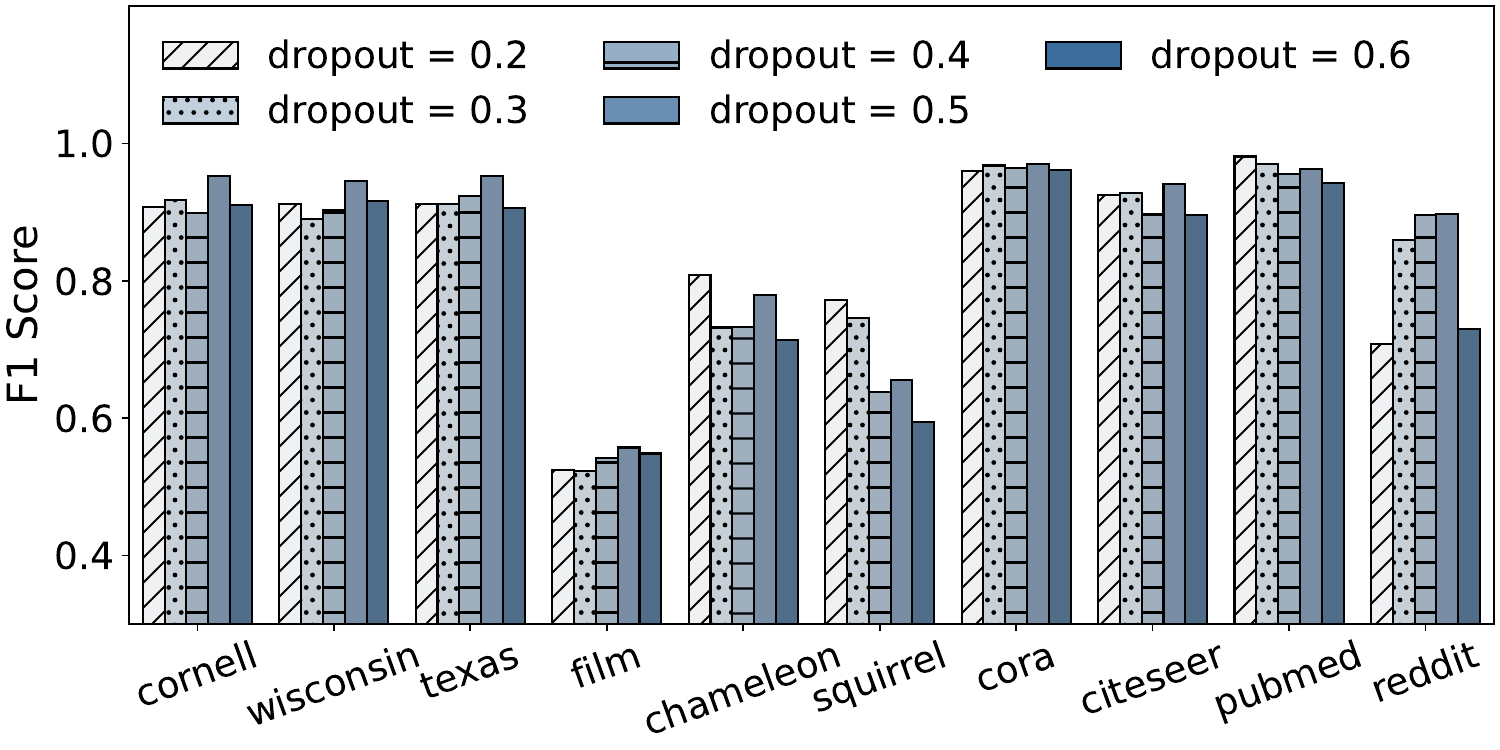}
        \vspace{-6mm}
        \caption{F1-Score by varying dropout ratio}
        \vspace{-6mm}
        \label{fig:chart6}
    \end{subfigure}
        \vspace{-2mm}
    \caption{\myblue{Hyper-parameter analysis}}
        \vspace{-6mm}
    \label{fig:hyper_analysis}
\end{figure*}

\subsection{Efficiency of AdaptCS}
\autoref{fig:efficiency} reports the training time of all learning-based models and the query efficiency of all search algorithms. 
ACS shares the same offline time as SCS, so it is omitted for clarity.
\myblue{The one-time offline cost (e.g., model training or preprocessing) and the online per-query latency are reported separately: Fig.~\ref{fig:efficiency}(a) measures the offline training time for learning-based models, while Fig.~\ref{fig:efficiency}(b) measures the online query time per query after the offline stage is completed.}


\myparagraph{Training phase}
\myblue{For a fair comparison, we report the offline training time of all learning-based baselines under the same light heterophily extension (ACM).}
SVD compression eliminates OOM and shortens runtime.
The numbers reported for AdaptCS training time included the one-off SVD pre-computation cost.
Overall, AdaptCS-II is the only model that fits into a single GPU on the 110-million-edge {Reddit} graph, whereas every other GNN variant crashes with OOM.
On small graphs, AdaptCS-II has comparable training time to AdaptCS-I, while on large graphs, it is faster due to the SVD approximation.
Compared to baselines, both AdaptCS variants demonstrate better training efficiency.

\myparagraph{Query phase}
\myblue{The considered training extensions (e.g., ALT/ACM/TANH) only affect the offline training stage and do not change the online search procedure; therefore, the per-query efficiency reported below is identical across these extensions for the same model.}
AdaptCS-SCS achieves sub-millisecond latency on most graphs,
          while AdaptCS-ACS achieves the best efficiency and stays below 0.03s on large graphs.
          \myblue{Compared with QDGNN or COCLEP, which require several seconds of
          forward propagation per query on heterophilic graphs such as
          {Film}, both AdaptCS objectives achieve up to 2 orders of magnitude acceleration over the strongest ML-based CS baselines.}
          Classical heuristics (clique, $k$-truss, $k$-core) return results in microseconds on small graphs, and ACS matches their efficiency on large graphs.
          
To sum up, the SVD-optimized {AdaptCS-II} not only attains the best accuracy but also solves the memory bottleneck: it is the only method that completes training on all datasets, and its query latency remains competitive with heuristics while outperforming all learning-based baselines by large margins.

\vspace{-1mm}
\subsection{Ablation Study}
\autoref{fig:ablation} illustrates the impact of five key components in {AdaptCS}: adaptive masking, the Signed Community Search (SCS), weight renormalization, MLP compared to the per-class bank fusion, and the random walk renormalized matrix.  
We report five variants—{w/o mask}, {w/o scs}, {w/o norm}, {w/o mlp}, {w/o sym}, and the {full model} (AdaptCS-II + ACS).

\myparagraph{Adaptive masking}
Replacing the adaptive masking with hard masking ({w/o mask}) results in the significant drop across most datasets, confirming that isolating exact-$k$ neighbors while retaining triangle-rich edges is crucial for model performance.

\myparagraph{Signed community search}
Substituting SCS with a plain BFS expansion ({w/o scs}) lowers F1 by up to 7\% on Chameleon.  
This improvement is primarily due to searching on the signed positive graph constructed from learned embeddings, which provides meaningful connectivity for community retrieval.

\myparagraph{Weight renormalization}
Eliminating the normalizer degrades performance across all datasets, indicating that rescaling hop-wise signals stabilizes training and prevents oversmoothing.

\myparagraph{MLP vs. attention}
Our framework supports both MLP-based fusion and a per-class attention bank for hop channel mixing.
\myblue{Overall, both fusion strategies are effective, while the attention-based fusion is stronger on most benchmarks; the MLP variant remains competitive and achieves comparable results on several datasets.
We attribute the advantage of attention to its node- and class-adaptive weighting of hop channels, which better aligns the fused representation with query-relevant semantics, especially under heterophily.}


\myparagraph{Random walk renormalized matrix}
Replacing the symmetric normalization with a random-walk normalization ({w/o sym}) yields comparable performance across all datasets, showing only marginal differences on small and medium graphs.
This indicates that both normalization schemes capture similar spectral properties, and the choice between them has a limited impact on overall community search performance.

\myparagraph{Full model}
The complete configuration (AdaptCS-II + ACS) achieves the highest F1 on most datasets and is the only variant that solves all benchmarks, including the 110-million-edge {Reddit}.  
These results demonstrate that each component contributes to the effectiveness and efficiency of {AdaptCS}.

\subsection{Hyper-parameter Analysis}
\autoref{fig:hyper_analysis} presents the sensitivity analysis of six key hyper-parameters in AdaptCS-II.
In each plot, we vary one parameter while keeping all others fixed at their default values, and report the average F1-score over 50 query nodes.

\myparagraph{Similarity threshold \(\boldsymbol{\tau}\)  — \autoref{fig:chart1}}  
Similarity threshold serves the purpose of transforming the cosine similarity of the learned embeddings into
a signed graph; an edge \((u,v)\) is labeled positive if \(s_{uv}\geq\tau\) and negative if \(s_{uv}<\tau\).
We test the threshold range of \(\tau\!\in\!\{0.2,0.5,0.6,0.8,0.9\}\).
From the study, we find that scores remain stable between
\(0.5\!\le\!\tau\!\le\!0.8\) and peak at \(\tau\!=\!0.9\).
\myblue{
A very small threshold (\(\tau\!=\!0.2\)) falsely turns many weakly
related pairs into positive edges, whereas a strict threshold (\(\tau\!=\!0.9\)) assign more weights to nodes that demonstrate high query similarity globally.  
We retain \(\tau\!=\!0.9\) as the default setting for experiments.
Since similarity score distributions can vary across graphs, \(\tau\) can also be set in a graph-adaptive manner, e.g., by choosing \(\tau\) as the \(p\)-quantile of sampled edge similarities. This provides a simple option to stabilize the positive-edge ratio on unseen graphs.
}

\myparagraph{SVD rank \(\boldsymbol{r}\) — \autoref{fig:chart2}}  
The SVD rank \(r\) determines the number of leading spectral components used in the low-rank approximation. 
Notably, a clear improvement is observed at \(r = 100\), where performance peaks or reaches its best on several datasets. Based on this observation, we set \(r = 100\) as the default, with \(r = 128\) used only for the largest graph ({Reddit}) to ensure maximum coverage.

\myparagraph{Hop number \(\boldsymbol{k}\) — \autoref{fig:chart3}}  
The hop count decides how many exact-hop channels are aggregated, balancing receptive fields against noise.  
We vary \(k\) between 3 and 9.  
\myblue{Performance increases as $k$ grows up to $5$, while further hops provide diminishing gains and can hurt accuracy on sparse graphs such as {Chameleon}. Since long-range propagation tends to introduce weakly related nodes, we set $k=5$ for all experiments.}

\myparagraph{Hidden dimension \(\boldsymbol{h}\) — \autoref{fig:chart4}}  
The hidden size determines the capacity of the projection MLP that follows channel mixing.  
Tested values range from 64 to 1024.  
F1 climbs steadily up to \(h\!=\!512\); larger widths contribute may over-fit small graphs, while increasing training time. 
Hence, we adopt \(h\!=\!512\) as a balanced setting in the experiments.

\myparagraph{Top factor \(\boldsymbol{\alpha}\) — \autoref{fig:chart5}}  
The top factor $\alpha$ in ACS determines the number of most similar nodes included in the candidate pool. As shown in \autoref{fig:chart5}, setting $\alpha=2$ consistently achieves the best F1-score across most datasets. Increasing $\alpha$ beyond 2 introduces more unrelated nodes into the candidate set, which not only degrades retrieval quality—especially on datasets like {Chameleon} and {Squirrel}—but also slows down the search. Therefore, we use $\alpha=2$ by default to maximize both accuracy and efficiency.

\myparagraph{Dropout ratio — \autoref{fig:chart6}}  
Dropout serves as a regularization tool for the embedding MLP.  
We evaluate dropout ratios ranging from 0.2 to 0.6.  
The results remain stable, with peak performance observed at dropout values of 0.2 and 0.5.  
Based on this, we adopt $\text{dropout}=0.5$ as the default setting.

Across all six hyperparameters, \textsc{AdaptCS} maintains strong performance over wide ranges, confirming that the defaults \(\{\tau=0.9,\,r=100,\,k=5,\,h=512,\,\alpha=2,\,\text{dropout}=0.5\}\) chosen offer a balance between effectiveness and efficiency.

\section{Case Study} \label{case}
\begin{figure}[t]
  \centering
  \includegraphics[width=\linewidth]{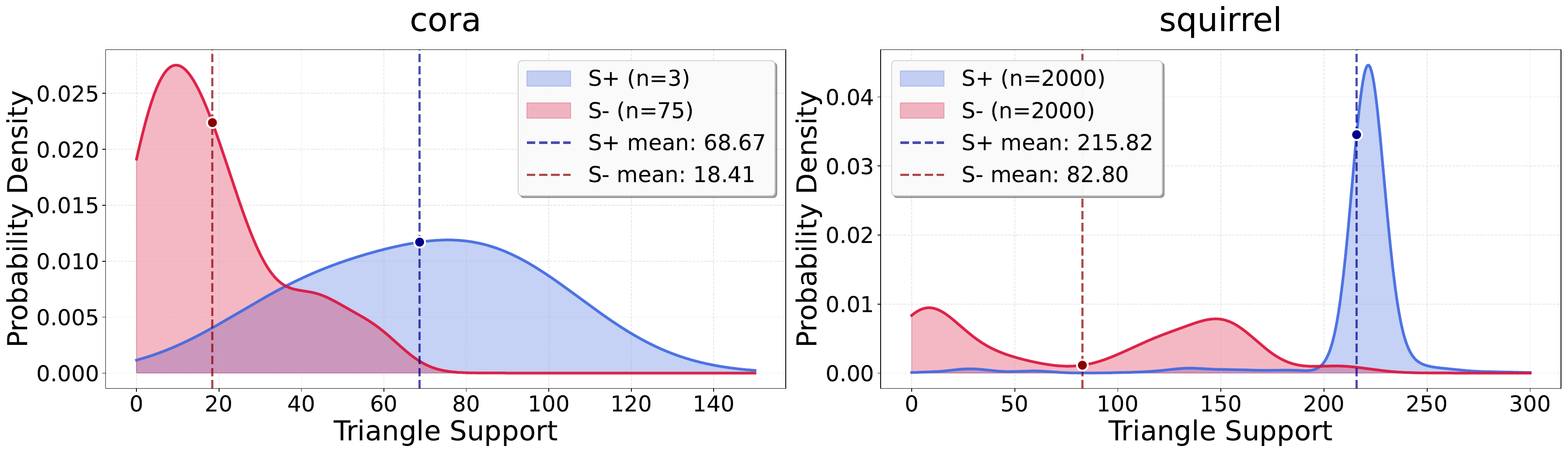}
        \vspace{-5mm}
  \caption{\myblue{Case Study: Distribution of Triangle Support}}
  \label{fig:case}
\end{figure}

\myblue{
\autoref{fig:case} compares the triangle support of direct edges that are retained in 2-hop as strengthened by the adaptive mask ($S^+$) against those that are not ($S^-$). 
Across datasets, the $S^+$ distribution shifts to the right, evidencing heavier tails and larger means (e.g., Cora: $68.7$ vs.\ $18.4$; Squirrel: $215.8$ vs.\ $82.8$), which indicates that adaptively retained edges participate in more triangles than their non-retained counterparts.

\myparagraph{Implication}
Edges with higher triangle support anchor dense subgraphs: multiple common neighbors provide more two-hop paths that (i) concentrate local cohesiveness within the same semantic group, making endpoints more likely to belong to the same community, and (ii) improve robustness under heterophily, since cross-class or noisy links are diluted by several consistent neighbors. 
Hence, adaptive masking preserves triangle-rich edges that benefit community identification.
}

\vspace{-1mm}
\section{Conclusions} \label{Conclusions}
In this study, we introduce AdaptCS, a novel community search framework adaptively designed for both homophilic and heterophilic graphs. 
By incorporating distance awareness through distinctive-hop aggregation and frequency awareness via low- and high-pass filtering, the AdaptCS encoder produces embeddings that remain discriminative even when neighboring nodes are largely dissimilar.
A scalable low-rank SVD optimization further removes the memory bottleneck of high-order adjacency computation, enabling efficient training on graphs with over one hundred million edges.
Leveraging these embeddings, the Adaptive Community Score (ACS) balances embedding similarity and topological relations, supporting accurate and efficient query-time retrieval.
Extensive experiments on both heterophilic and homophilic benchmarks demonstrate that AdaptCS consistently outperforms state-of-the-art baselines, maintains robustness under varying degrees of heterophily, and achieves substantial gains in efficiency.

\section{Statement on AI-assisted tools}
Portions of this work were assisted by AI tools for language proofreading, formatting refinement, and minor code debugging. 
All conceptual development, experimental design, analysis, and writing decisions were made by the authors.

\bibliographystyle{IEEEtran}
\bibliography{ref}

\end{document}